\documentclass[reqno]{amsart}
\usepackage{amsfonts,amssymb,anysize}
\usepackage{multicol, color}
\usepackage{graphicx}
\usepackage{amsmath, amsthm}

\newcommand{\rem}[1]{}

\newtheorem{lemma}{Lemma}

\newtheorem{proposition}[lemma]{Proposition}

\theoremstyle{remark}

\newcommand*  {\R} {{\mathbb R}}

\def\aa{\alpha}
\def\a2{\alpha^2}

\def\dd{\delta}

\def\ss{\sigma}

\def\gd{\nabla}
\def\lp{\triangle}

\begin{document}
\DeclareGraphicsExtensions{.pdf, .gif, .jpg, .eps}
\title[Modulation Approximation]
{Modulation theory for self-focusing in the nonlinear Schr\"{o}dinger-Helmholtz equation}
\date{February 8, 2009.\\ {\bf To appear in:} {\it
Numerical Functional Analysis and Optimization} }

\author[Y.Cao]{Yanping Cao}
\address[Y.Cao]
{Department of Mathematics\\
University of California\\
Irvine, CA 92697-3875,USA}
\email{ycao@math.uci.edu}

\author[Z. H. Musslimani]{Ziad H. Musslimani}
\address[Z. H. Musslimani]
{Department of Mathematics\\
Florida State University\\
Tallahassee FL 32306, USA}
\email{musliman@mail.math.fsu.edu}

\author[E.S. Titi]{Edriss S. Titi}
\address[E.S. Titi]
{Department of Mathematics \\
and  Department of Mechanical and  Aerospace Engineering \\
University of California \\
Irvine, CA  92697-3875, USA \\
{\bf ALSO}  \\
Department of Computer Science and Applied Mathematics \\
Weizmann Institute of Science  \\
Rehovot 76100, Israel}
\email{etiti@math.uci.edu and edriss.titi@weizmann.ac.il}

\begin{abstract}
The nonlinear Schr\"{o}dinger-Helmholtz (SH) equation in $N$ space dimensions with
$2\sigma$ nonlinear power was proposed as a regularization of the classical nonlinear Schr\"{o}dinger (NLS) equation. It was shown that the SH equation has a larger regime
($1\le\sigma<\frac{4}{N}$) of global existence and uniqueness of solutions compared to that of the classical NLS ($0<\sigma<\frac{2}{N}$). In the limiting case where the Schr\"{o}dinger-Helmholtz equation is viewed as a perturbed system of the classical NLS equation, we apply modulation theory to the classical critical case ($\sigma=1,\:N=2$) and show that the regularization prevents the formation of singularities of the NLS equation. Our theoretical results are supported by numerical simulations.
\end{abstract}

\maketitle

{\bf MSC Classification}: 35Q40, 35Q55, 78A60 \\

{\bf Keywords}:  Perturbed critical nonlinear Schr\"{o}dinger equation, Hamiltonian, Regularization of the nonlinear Schr\"{o}dinger equation, Schr\"{o}dinger-Helmholtz Equation,  Schr\"{o}dinger-Newton equation, Modulation theory.
\section{Introduction}\label{SEC-intro}
The nonlinear Schr\"{o}dinger-Helmholtz (SH) system is given as below
\begin{eqnarray}
&& iv_t+\lp v+u|v|^{\ss-1} v=0 , \qquad \qquad t>0,\:x\in \R^N,       \nonumber         \\
&& u-\alpha^2 \lp u=|v|^{\ss+1} ,                                                         \label{SH}      \\
&& v(0)=v_0.                                                                           \nonumber
\end{eqnarray}
where $\ss\ge1$ and, for simplicity, $\alpha>0$. This system has been proposed \cite{CMT} as a regularization for the classical nonlinear
Schr\"{o}dinger (NLS) equation:
\begin{eqnarray}
&& iu_t+\lp u+|u|^{2\ss} u=0,\qquad \qquad t>0,\: x \in \R^N, \nonumber \\
&&u(0)=u_0.          \label{NLS}
\end{eqnarray}
In \cite{CMT} we showed global existence of solution of the Cauchy problem (\ref{SH}) for
$1\le \ss <3$ when $N=1$ and $1\le \ss < \frac{4}{N}$ when $N>1$.
It is well-known that the classical NLS has global solution for $0\le \ss <\frac{2}{N}$ in any
dimension $N\ge1$ and there is finite time blow up in the critical case $\ss=\frac{2}{N}$(see e.g.,
\cite{Cazenave}, \cite{Velo}, \cite{Glassey}, \cite{Kato}, \cite{Sulem}, \cite{Weinstein} and references
therein). So we regard the SH system (\ref{SH}) as a regularization system for the NLS (\ref{NLS}) since the former system has larger regime of global existence for the parameter $\ss$, which contains the values for which the NLS (\ref{NLS}) blows up.  Note that (\ref{SH}) is a Hamiltonian system with the corresponding Hamiltonian
$$\mathcal{H}(v)=\int_{\R^N} \left( |\gd v(x,t)|^2-\frac{u(x,t)|v(x,t)|^{\ss+1}}{\ss+1} \right) \:dx$$
and can be obtained formally by the variational principle
$$i\frac{\partial v}{\partial t} =\frac{\dd \mathcal{H} (v) }{\dd v^*},$$
where $v^*$ denotes the complex conjugate of $v$.
Let us rewrite system (\ref{SH}) as
\begin{equation}
\hskip-.8in
iv_t +\lp v +|v|^{2\ss} v + \alpha^2( \lp u)|v|^{\ss-1} v=0,      \label{pert-SNG}
\end{equation}
where $u=(I-\alpha^2 \lp)^{-1} (|v|^{\ss+1})$.  Observe that when the parameter $\alpha$ goes to zero, one can regard system (\ref{SH}) (or \ref{pert-SNG}) as a formal perturbation of the classical NLS.
There has been a lot of work on perturbed NLS in the critical case $\ss=\frac{2}{N}$ ( see, e.g.,
\cite{FibichDamp}, \cite{FIP}, \cite{FibichLevy}, and references therein). In this paper, we
will apply modulation theory (see e.g., \cite{FP1}, \cite{FP}, \cite{Landman} and \cite{Malkin} for references about
modulation theory) to the classical critical case $\ss=1, N=2 $ in order to shed more light on the nature
of the effect of the regularization in preventing the blow up. In this case the classical NLS blows up for
certain initial data, however, the SH system has global solution with the regularization parameter $\aa>0$. Indeed, modulation theory tries to explain the role of the regularization in preventing the formation of singularity near the critical values of the initial data which blow up in the classical case.
Intuitively speaking, the basic idea behind modulation theory is that the energy near singularity is equal to the power of the Townes soliton [see Eq.~(\ref{soliton}) below], and the profiles of
the solutions are asymptotic to some rescaled profiles of the Townes soliton.
With modulation theory, one can reduce the perturbed system ({\ref{pert-SNG}) into a simpler system
of ordinary differential equations that do not depend on the spatial variables, and they are supposed to
be easier to analyze both analytically and
numerically. \\\\
In this paper, we will study how the parameter $\alpha$ prevents the singularity formation in the 2D critical NLS. The work here will follow the study we initiated in \cite{CMT}. In particular, some of
the statements have already been mentioned there. For the sake of completeness, we will restate
some important theorems and propositions.
\section{Modulation theory}  \label{SEC-modulation}
First we review some main results on modulation theory for the unperturbed critical NLS following \cite{FP}. As stated in \cite{FP}, most of the results presented in this section are formal and have not been made rigorous at present. We emphasize here that we consider the case $\ss=1,\: N=2$ in order
to see how the regularization prevents singularity formation.\\

In the case of self-focusing the amount of power which goes into the singularity is equal to the critical power
$N_c=\|R\|_2^2=\int_0^\infty R^2(r) r\: dr$ where $R$, the Townes soliton, is the solution of the following equation with minimal $L^2-$norm, which is positive and radially symmetric
\begin{equation}
\hskip-.8in
\lp R-R+R^3=0, \quad R'(0)=0, \quad \underset{r
\rightarrow +\infty}{\lim}R(r)=0, \quad \lp=\frac{\partial^2}{\partial r^2}+\frac{1}{r}\frac{\partial}{\partial r}. \label{soliton}
\end{equation}
Close to the stage of self-focusing, the solution of (\ref{NLS}) separates into two components as it propagates,
\begin{equation}
\hskip-.8in
v=v_s+v_{\mbox{back}}
\end{equation}
where $v_s$ is the high intensity inner core of the beam which self-focuses toward its center axis
and $v_{\mbox{back}}$ is the low intensity outer part which propagates forward following the usual linear propagation mode. Close enough to the singularity, $v_s$ has only small excess power above the critical one and approches the radially symmetric asymptotic profile:
\begin{equation}
\hskip-.8in
v_s(x,t)=\frac{1}{L(t)}V(\tau, \xi) \exp \left( i (\tau+\frac{L_t}{L}\frac{r^2}{4} ) \right), \qquad  argV(\tau,0)=0,
\end{equation}
where $L(t)$ is a yet undetermined function that is used to rescale the variables $\xi=(\xi_1,\xi_2),\: x=(x_1,x_2)$ with
\begin{equation*}
\hskip-.8in
\xi=\frac{x}{L}, \: \frac{d\tau}{dt}=\frac{1}{L^2}.
\end{equation*}
Then the reduced system for unperturbed critical NLS is
\begin{eqnarray}
\hskip-.8in
&& L_{tt}=-\frac{\beta}{L^3},       \nonumber         \\
\hskip-.8in
&& \beta_t=-\frac{ e^{-\frac{\pi}{\sqrt{\beta}}}}{L^2},   \label{unpert-reduced}
\end{eqnarray}
where $L(t)$ is the scaling factor and $\beta$ is proportional to the excess of the power near
singularity: $\beta=M(N-N_c)$ for constant $M=\frac{1}{4} \int_0^{\infty}R^2(\rho) \rho^3 \: d\rho \approx 0.55$. We emphasize here that in the case of self-focusing of the original system, both the scaling factor $L$ and the excess energy $\beta$ will approach zero.\\\\
Next we review some results on modulation theory for the perturbed critical NSL \cite{FP}. For a general perturbed critical NLS of the form
\begin{equation}
\hskip-.8in
iv_t+ \lp v + |v|^2 v+ \epsilon F( v, v_t, \gd v, \cdot \cdot \cdot )=0, \quad |\epsilon| \ll 1,  \label{PCNLS}
\end{equation}
where $F$ is an even function in $x$, modulation theory is valid when the following three conditions hold.\\\\
{\bf{Condition 1.}} The focusing part of the solution is close to the asymptotic profile
\begin{equation}
\hskip-.8in
v_s(t,x) \sim \frac{1}{L(t)} V(\tau,\xi ) \exp [ i \tau(t)+i\frac{L_t}{L}\frac{r^2}{4}],
\label{condition1}
\end{equation}
where$$ \xi=\frac{x}{L}\;,\;\;\;\;\;  r^2=x_1^2+x_2^2\;,\;\;\;\;\; \frac{d \tau}{dt}=\frac{1}{L(t)^2}$$
and $V=R+\mathcal{O}(\beta, \epsilon), \: \beta=-L^3L_{tt}$ and $R$ is the Townes soliton given in (\ref{soliton}).\\\\
{{\bf{Condition 2.}} The power is close to critical
\begin{equation}
\hskip-.8in
| \frac{1}{2\pi} \int |v_s (t,x_1,x_2)|^2\: dx_1dx_2 -N_c| \ll 1,  \label{condition2}
\end{equation}
or, equivalently,
\begin{equation}
\hskip-.8in
| \beta(t)| \ll 1,    \label{condition22}
\end{equation}
where $N_c=\frac{1}{2\pi} \int_{\R^2} R(x_1,x_2)^2 \:dx_1\:dx_2=\|R\|_2^2$ is the threshold energy of blowup.\\\\
{\bf{Condition 3.}} The perturbation $\epsilon F$ is small in comparison to the other terms, i.e.,
\begin{equation}
\hskip-.8in
| \epsilon F| \ll | \lp v|, \: | \epsilon F| \ll |v|^3.
\label{condition3}
\end{equation}\\
The following proposition is given in \cite{FP}.
\begin{proposition} If conditions 1-3 hold, self-focusing in the perturbed critical NLS
(\ref{PCNLS})  is given to leading order by the reduced system
\begin{equation}
\hskip-.8in
\beta_t+ \frac{e^{-\frac{\pi}{\sqrt{\beta} } }}{L^2}=\frac{\epsilon}{2M}(f_1)_t-\frac{2\epsilon}{M}f_2\;,\;\;\;\;\;
L_{tt}=-\frac{\beta}{L^3}.                \label{reduced-system}
\end{equation}
The auxiliary functions $f_1,f_2$ are given by
\begin{eqnarray}
\hskip-.8in
&&f_1(t)=2L(t)Re\left[  \frac{1}{2\pi} \int_{\R^2} F( \psi_R) \exp (-iS)[R(\rho)+\rho \gd R(\rho) ] dx_1dx_2 \right],    \label{f-1}   \\
\hskip-.8in
&&f_2(t)=Im \left[ \frac{1}{2\pi} \int_{\R^2} \psi_R^*F(\psi_R) dx_1dx_2 \right], \label{f-2}
\end{eqnarray}
where
$$\psi_R=\frac{1}{L}R(\rho)\exp (iS),$$
and $R$ is the Townes soliton given in (\ref{soliton}), $\: \rho=\frac{r}{L}, \: S=\tau(t)+\frac{L_t}{L}\frac{r^2}{4}, \:
\frac{d\tau}{dt}=\frac{1}{L^2}, \: M=\frac{1}{4} \int_0^{\infty} R(\rho)^2 \rho^3 \: d\rho\approx 0.55$.\\

Furthermore, if $F$ is a conservative perturbation, i.e.,
\begin{equation}
Im\int_{\R^2} v^* F(v) dx_1dx_2 =0,    \label{conservative}
\end{equation}
then $f_2$=0. Since $\beta\ll 1,\: e^{-\frac{\pi}{\sqrt{\beta}}} \ll \beta$, taking the leading order by neglecting the exponential term in the first equation of (\ref{reduced-system}), we further reduce the
system (\ref{reduced-system}) into the following system
\begin{equation}
\hskip-.8in
-L^3L_{tt}=\beta=\beta_0+\frac{\epsilon}{2M}f_1\;,\;\;\;\;\; \beta_0=\beta(0)-\frac{\epsilon}{2M}f_1(0), \label{reduced-reduced}
\end{equation}
where $\beta_0$ is independent of $t$.
\end{proposition}
In general, at the onset of self-focusing only condition 3 holds. Therefore, if the power is above
$N_c$ the solution will initially self-focus as in the unperturbed critical NLS. As a result, near the time of blowup in the absence of the perturbation, conditions $1-2$ will also be satisfied.

It is worth pointing out that, as studied in \cite{FP}, various conservative perturbations of the critical NLS equation, for instance, self-focusing in fiber arrays (see \cite{AALRT}, \cite{AART}, \cite{AALRT2},
\cite{AALRT3}, \cite{LST}, \cite{WY} and references therein) and small dispersive fifth-power nonlinear perturbation to the classical NLS \cite{Malkin}, have a generic form
\begin{equation}
\hskip-.8in
f_1 \sim-\frac{C}{L^2},\: C=\mbox{constant},         \label{f1-generic}
\end{equation}
which results in a canonical focusing-defocusing oscillation. \\\\

Next, we shall derive the reduced equations (\ref{reduced-reduced}) that corresponds to the nonlinear Schr\"{o}dinger-Helmholtz regularization system in the critical case $\ss=1,\:N=2$ (the reason for this restriction is that it allows us to compare with numerical simulations).  In this case, Eq.~(\ref{pert-SNG}) reads
\begin{equation}
\hskip-.8in
iv_t+\lp v+|v|^2v+\alpha^2v\lp u =0\;. \label{critical-perturbed}
\end{equation}
Comparing Eq.~(\ref{critical-perturbed}) with (\ref{PCNLS}) we have $\epsilon =\alpha^2$ and
\begin{equation}
\hskip-.8in
F(v)=v\lp u\;,\;\;\;\;\;\;\;\;\;\;u-\alpha^2 \lp u =|v|^2. \label{Fv}
\end{equation}
We shall assume that the system (\ref{critical-perturbed}) satisfies all three conditions, (\ref{condition1}), (\ref{condition2}) and (\ref{condition3}). Since
\begin{equation}
\hskip-.8in
\psi_R(x)=\frac{1}{L}R(\frac{x}{L})\exp(iS),
\end{equation}
we have
\begin{equation}
\hskip-.8in
F(\psi_R)(x)=\psi_R (x)\lp u_R  \label{F},
\end{equation}
where $u_R$ satisfies
\begin{equation}
u_R (x)- \alpha^2 \lp u_R(x)= |\psi_R(x)|^2=|\frac{1}{L} R(\frac{x}{L} ) \exp(iS)|^2=\frac{1}{L^2}| R(\frac{x}{L})|^2 \:.   \label{laplace}
\end{equation}
For a given function $g$, solution to the equation
\begin{equation}
(I-\alpha^2 \lp) u(x)= g( \frac{x}{L}), \qquad \qquad x\in \R^2
\end{equation}
is given by
$$u(x)=(B_{\frac{\alpha}{L}} \ast g) (\frac{x}{L})\;,$$
where $B_{\frac{\alpha}{L}}$ is the modified Bessel potential or the Green function corresponding to the Helmholtz operator (see e.g., \cite{Evans} for reference on Bessel potential)
\begin{equation}
B_{\frac{\alpha}{L}}(x)=\frac{1}{2\alpha^2} \int_0^{\infty}
\frac{ e^{-s} e^{-\frac{|x|^2}{4s (\alpha/L)^2} }} { s^{N/2} } \: ds.
\end{equation}
If we let $g(\cdot)=\frac{1}{L^2} R(\cdot)^2$ then we can write the solution to Eq.~(\ref{laplace}) as
\begin{eqnarray}
u_R(x)
&= &(B_{\frac{\alpha}{L}} \ast \frac{1}{L^2}R^2) (\frac{x}{L})=\frac{1}{L^2} (B_{\frac{\alpha}{L}} \ast
R^2 ) (\frac{x}{L})              \\
&=&\frac{1}{L^2} \frac{1}{2\alpha^2} \int_{\R^2} \left(     \int_0^{\infty}
\frac{ e^{-s} e^{-\frac{|x/L-y|^2}{4s (\alpha/L)^2} }} { s } \: ds      \right) R^2(y) dy_1dy_2 ,     \label{ur}
\end{eqnarray}
where $y=(y_1,y_2)$. Substituting the above into (\ref{F}) and using (\ref{laplace}), we get
\begin{eqnarray}
F(\psi_R)(x)
&=&\psi_R(x)\lp u_R(x)=\frac{1}{\alpha^2} ( u_R(x)- |\psi_R(x)|^2)  \psi_R(x)          \\
&=&\frac{1}{\alpha^2L}R(\frac{x}{L})\left[u_R(x)-\frac{1}{L^2}R^2(\frac{x}{L})\right]e^{iS}.
\end{eqnarray}\\\\
Now we can calculate the term $f_1$
\begin{eqnarray*}
f_1
&=& \frac{L}{\pi} Re \int_{\R^2} F(\psi_R(x)) \exp(-iS) ( R(\rho)+ \rho R_{\rho} ) dx_1 dx_2     \\
&=&\frac{L}{\pi} \int_{\R^2} \frac{1}{\alpha^2L} R(\rho)[ u_R(x)- \frac{1}{L^2} R^2(\rho) ] [ R(\rho)+ \rho R_{\rho} ] \: dx_1 dx_2          \\
     &=&\frac{1}{\pi \alpha^2} \int_{\R^2} u_R(x) R(\rho) [R(\rho)+ \rho R_{\rho}] dx_1dx_2- \frac{1}{\pi \alpha^2} \frac{1}{L^2} \int_{\R^2} R^3(\rho)[R(\rho)+ \rho R_{\rho} ] dx_1 dx_2       \\
     &=&J_1-J_2.
\end{eqnarray*}
It is easy to see that the second integral $J_2$ is a constant that does not depend on $L$. Indeed, applying change of variables: $\xi_1=\frac{x_1}{L},\xi_2=\frac{x_2}{L}$, since $\rho=\frac{r}{L}=
\frac{\sqrt{x_1^2+x_2^2}}{L}=\sqrt{\xi_1^2+\xi_2^2}$, we get
\begin{equation}
\hskip-.8in
J_2=\frac{1}{\pi \alpha^2} \int_{\R^2}R^3(\rho)( R(\rho)+\rho R_{\rho}(\rho)) \:d\xi_1 d\xi_2=c_0, \label{I-2}
\end{equation}
where $c_0$ is a constant that does not depend on $L$.\\

Next, let us look at the first integral $J_1$. Plugging the result of (\ref{ur}) into $J_1$, we get
\begin{eqnarray}
&&J_1= \frac{1}{\pi \alpha^2} \int_{\R^2} u_R(x) R(\rho) ( R(\rho)+\rho R_{\rho} ) dx_1dx_2      \\
&& =\frac{1}{\pi \alpha^2} \int_{\R^2 } \frac{1}{L^2} \frac{1}{2\alpha^2} \int_{\R^2} \left( \int_0^{\infty}
\frac{e^{-s}e^{-\frac{|x/L-y|^2}{4s(\alpha/L)^2}} }{s} ds \right) R^2(y) \:dy_1dy_2 \: R(\rho)( R(\rho)+ \rho R_{\rho} ) dx_1dx_2   \\
&& =\frac{1}{2\pi} \frac{1}{\alpha^4} \frac{1}{L^2} \int_{\R^2} \int_{\R^2} \int_0^{\infty}
\frac{e^{-s} e^{-\frac{|x/L-y|^2}{4s(\alpha/L)^2}}}{s} \: ds \: R^2( y) \: dy_1dy_2\:
R(\rho) (R(\rho)+\rho R_{\rho})  d x_1dx_2 .
\end{eqnarray}
Change of variables: $ \xi=(\xi_1,\xi_2)=(\frac{x_1}{L},\frac{x_2}{L})$, then we will have
\begin{equation}
\hskip-.8in
J_1=\frac{1}{2\pi}\frac{1}{\alpha^4} \int_{\R^2} \int_{\R^2} \int_0^{\infty}
\frac{e^{-s} e^{-\frac{|\xi-y|^2}{4s(\alpha/L)^2}}}{s} \: ds \: R^2( y) \: dy_1dy_2 \:
R(\rho) (R(\rho)+\rho R_{\rho})  d\xi_1 \: d\xi_2.
\end{equation}
So $f_1$ can be written as
\begin{equation}
\hskip-.8in
f_1=\frac{1}{2\pi}\frac{1}{\alpha^4} \int_{\R^2} \int_{\R^2} \int_0^{\infty}
\frac{e^{-s} e^{-\frac{|\xi-y|^2}{4s(\alpha/L)^2}}}{s} \: ds \: R^2( y) \: dy_1dy_2 \:
R(\rho) (R(\rho)+\rho R_{\rho})  d\xi_1 \: d\xi_2 -c_0,                                                      \label{f1-final}
\end{equation}
where $c_0$ is a constant given in (\ref{I-2}).\\

Plugging into (\ref{reduced-reduced})  we have the reduced system for the Schr\"{o}dinger-Helmholtz
system (\ref{SH})
\begin{equation}
\hskip-.8in
-L^3L_{tt}=\beta_0+\frac{\alpha^2}{M}f_1,\: \beta_0=\beta(0)-\frac{\alpha^2}{2M}f_1(0) \label{SH-reduced}
\end{equation}
where $f_1$ is given in (\ref{f1-final}). \\

Now one needs to study the ordinary differential equation (\ref{SH-reduced})  with $f_1$ given in (\ref{f1-final}).
The explicit form of the function $f_1$ is much more complicated when compared to the generic form of (\ref{f1-generic}) due to the nonlocal nature of our perturbation term (\ref{Fv}).
The idea is to show that for $L$ small this additional term on the right-hand side of the first equation
of (\ref{SH-reduced}) will prevent the singularity formation, i.e., prevents $L$ from tending to zero as time evolves. In the next section, we will investigate the ODE system (\ref{SH-reduced}) by approximating
the function $f_1$.
\section{A simplified Reduced system}\label{SEC-appro}
Now, without solving $u_R(x)$ explicitly as we did above, we will try to approximate
$u_R(x)$ by asymptotic expansion in terms of $\frac{\alpha}{L}$, for small values of $\frac{\alpha}{L}$,
and further approximate the function $f_1$.
\subsection{First order approximation}\label{first-order}
Recall that from (\ref{SH}) we have (when $\ss=1,\:N=2$)
\begin{equation*}
u-\alpha^2 \lp u=|v|^2
\end{equation*}
or we can write $u(x)=(I-\alpha^2 \lp )^{-1}|v(x)|^2$. When $\alpha$ is very small, we can
formally write $u(x)$ in the first leading order term:  $u(x)=|v(x)|^2+\mathcal{O} (\alpha^2) $.
Now for $\psi_R(x)=\frac{1}{L} R(\frac{x}{L}) \exp{(iS)}$, we can similarly write $u_R(x)$ as
\begin{eqnarray}
u_R(x)
&=&|\psi_R(x)|^2+\mathcal{O}((\frac{\alpha}{L})^2)         \nonumber   \\
&= &\frac{1}{L^2} R^2(\frac{x}{L})+\mathcal{O}((\frac{\alpha}{L})^2)   \label{first-order}
\end{eqnarray}
so we have
\begin{eqnarray}
F(\psi_R)(x)
&=&(\lp_x  u_R(x)) \psi_R(x)         \nonumber     \\
&\sim& ( \lp_x( |\psi_R(x)|^2))\psi_R(x)  \nonumber       \\
&\sim&  \frac{1}{L^2} ( \lp_x R^2(\frac{x}{L}) )\frac{1}{L}R(\frac{x}{L})\exp{(iS)}
\end{eqnarray}
Substituting this into the equation for $f_1$ (\ref{f-1}), we get
\begin{eqnarray}
f_1
&=& \frac{L}{\pi} Re \int_{\R^2} F( \psi_R(x)) \exp {(-iS)} (R(\rho)+ \rho R_{\rho} ) dx_1dx_2  \nonumber  \\
&\sim& \frac{L}{\pi} \int_{\R^2}   \frac{1}{L^2}( \lp_x R^2(\frac{x}{L}) )\frac{1}{L}R(\frac{x}{L})(R(\rho)+\rho R_{\rho}) dx_1dx_2
\end{eqnarray}
Next, we make change of variables:
\begin{equation*}
\xi=\frac{x}{L}, \qquad \qquad \xi=(\xi_1,\xi_2)
\end{equation*}
then for $\rho=|\xi|$, and by the chain rule, we get
\begin{eqnarray}
f_1
&\sim& \frac{1}{\pi} \frac{1}{L^2}\int_{\R^2} (\lp_{\xi} R^2(\rho) ) R(\rho) (R(\rho)+\rho R_{\rho} ) d\xi_1d\xi_2     \nonumber       \\
&\sim& -\frac{C_1}{L^2} =I_1      \label{I1}
\end{eqnarray}
where
\begin{eqnarray}
\hskip-.8in
C_1
&=& -  \frac{1}{\pi} \int_{\R^2} (\lp_{\xi} R^2)  R(R+\rho R_{\rho} )d\xi_1 d\xi_2    \nonumber  \\
\hskip-.8in
&=& 2 \int_0^{\infty} [ (R^2)_{\rho} ]^2 \rho d\rho>0,      \label{C1}
\end{eqnarray}
where the detail of the calculation of the above integral is presented in claim 1 of the appendix.\\

Plugging into (\ref{SH-reduced}), we have the leading order of the reduced system, which turns out to be of the generic form
\begin{equation}
\hskip-.8in
-L^3L_{tt}=\beta_0+\frac{\alpha^2}{2M} f_1   \label{SH-redu}
\end{equation}
with $f_1\sim -\frac{C_1}{L^2}$ and $\beta_0=\beta(0)+\frac{\alpha^2C_1}{2M}\frac{1}{L^2(0)}\ll1$ since $\beta(0)\ll1,\:
\frac{\alpha}{L}\ll1$.
Fibich and Papanicolaou \cite{FP} showed that there is no singularity in finite time with
this perturbation. In fact, substituting $f_1$ (\ref{I1}) into the above equation, we get
\begin{equation}
\hskip-.8in
-L^3L_{tt}=\beta_0-\frac{C_1}{2M}\frac{\alpha^2}{L^2} .  \label{L1}
\end{equation}
Write $y=L^2$, then $y$ satisfies the following equation:
\begin{equation}
\hskip-.8in
(y_t)^2 =4\beta_0-\frac{\alpha^2C_1}{M} \frac{1}{y}+4D_0y,  \label{y1}
\end{equation}
where $D_0$ is a constant satisfying
$D_0=L_t^2(0)-\frac{\beta_0}{L^2(0)}+\frac{\alpha^2 C_1}{4M} \frac{1}{L^4(0)}$. \\
Since
$\frac{\alpha^2 C_1}{M}>0$, $y$ can not go to zero in the above equation, i.e., $L$ can not
go to zero, which explains the prevention of the singularity formation, at this leading order
in the expansion.
\subsection{Second order approximation}\label{next-order}
In the previous subsection we use the asymptotic expansion by taking the first leading term as approximation for the solution $u$ of the Helmholtz equation $u-\alpha^2 \lp u=|v|^2$. One might naturally ask whether we will get better approximation and still have the
no blow up structure if we approximate the solution of the Helmholtz equation $u(x)$ by taking one more leading term.
To investigate this, we proceed similarly as
before: we can
formally write $u(x)$ in the first two leading terms: $u(x)=|v(x)|^2+\alpha^2 \lp |v(x)|^2 +\mathcal{O} (\alpha^4) $,
so we have
\begin{eqnarray}
u_R(x)
&=&|\psi_R(x)|^2+\alpha^2 \lp_x |\psi_R(x)|^2 +\mathcal{O}((\frac{\alpha}{L})^4)          \nonumber \\
&=&\frac{1}{L^2} R^2(\frac{x}{L})+\alpha^2 \lp_x ( \frac{1}{L^2} R^2(\frac{x}{L} )) +\mathcal{O}( (\frac{\alpha}{L})^4)   \nonumber  \\
&\sim& \frac{1}{L^2}R^2(\frac{x}{L}) +(\frac{\alpha}{L})^2 \lp_x R^2 (\frac{x}{L}) \label{second-order}
\end{eqnarray}
and then
\begin{eqnarray*}
F(\psi_R)(x)
&=&(\lp_x  u_R(x)) \psi_R(x)      \\
&\sim&\left( \lp_x ( \frac{1}{L^2} R^2(\frac{x}{L}) +(\frac{\alpha}{L})^2\lp_x R^2(\frac{x}{L}) ) \right) \frac{1}{L} R(\frac{x}{L}) \exp{(iS)}  \\
&\sim&\left(  \frac{1}{L^2}  \lp_x  R^2(\frac{x}{L}) +(\frac{\alpha}{L})^2 \lp_x^2 R^2(\frac{x}{L}) \right)\frac{1}{L}R(\frac{x}{L})\exp{(iS)}
\end{eqnarray*}
Substituting this into the equation of $f_1$ (\ref{f-1}), we get
\begin{eqnarray*}
f_1
&=& \frac{L}{\pi} Re \int_{\R^2} F( \psi_R(x)) \exp {(-iS)} (R(\rho)+ \rho R_{\rho} ) dx_1dx_2  \\
&\sim&\frac{L}{\pi}  \int_{\R^2} ( \frac{1}{L^2}  \lp_x  R^2(\frac{x}{L})
+(\frac{\alpha}{L})^2 \lp_x^2 R^2(\frac{x}{L} ))\frac{1}{L}R(\frac{x}{L})(R(\rho)+ \rho R_{\rho} ) dx_1dx_2  \\
&\sim& \frac{1}{\pi} \int_{\R^2} \frac{1}{L^2} (\lp_x R^2(\frac{x}{L})) R(\frac{x}{L})(R(\rho)+\rho R_{\rho}) dx_1 dx_2  \\
&+& \frac{1}{\pi} \int_{\R^2} (\frac{\alpha}{L})^2\lp_x^2 R^2(\frac{x}{L}) R(\frac{x}{L}) ( R(\rho)+\rho R_{\rho}) dx_1 dx_2     \\
&\sim&I_1+I_2
\end{eqnarray*}
Here the first integral $I_1$ is equal to $-\frac{C_1}{L^2}$ with $C_1$ in (\ref{C1}) as we calculated in
the first order expansion case. For the second integral $I_2$, after change of variables $\xi=\frac{x}{L}, \xi=(\xi_1,\xi_2)$ and $ \rho=|\xi|$,  it gives
\begin{eqnarray}
I_2
&=&\frac{1}{\pi}\frac{\alpha^2}{L^4} \int_{\R^2} (\lp_{\xi}^2(R^2))R(R+\rho R_{\rho})d\xi_1 d\xi_2 \nonumber \\
&=& \frac{\alpha^2 C_2}{L^4},   \label{I2}
\end{eqnarray}
where
\begin{eqnarray}
C_2
&=&\frac{1}{\pi}\int_{\R^2} (\lp_{\xi}^2(R^2))R(R+\rho R_{\rho}) d\xi_1 d\xi_2  \nonumber \\
&=& \frac{3}{2\pi}\int_{\R^2} ( \lp_{\xi}R^2 )^2 d\xi_1 d\xi_2>0,     \label{C2}
\end{eqnarray}
where the detailed calculation of the above integral is presented in claim 2 of the appendix.\\

Therefore, we obtain
\begin{equation}
f_1\sim-\frac{C_1}{L^2}+ \frac{ C_2\alpha^2}{L^4}
\end{equation}
with a correction next order term to the generic form (\ref{f1-generic}). It yields the reduced equation of the leading order \begin{equation}
-L^3L_{tt}=\beta_0+\frac{\alpha^2}{2M} f_1.  \label{corr}
\end{equation}
Substituting $f_1$ into the above equation (\ref{corr}), we get
\begin{equation}
\hskip-.8in
-L^3L_{tt}=\beta_0-\frac{C_1}{2M}(\frac{\alpha}{L})^2 +\frac{C_2}{2M}(\frac{\alpha}{L})^4.   \label{L2}
\end{equation}
Then $y=L^2$ satisfies the following equation:
\begin{equation}
(y_t)^2 =4\beta_0-\frac{\alpha^2C_1}{M} \frac{1}{y}+\frac{2}{3}\frac{ \alpha^4 C_2}{M}\frac{1}{y^2}+E_0y, \label{y2}
\end{equation}
where $E_0$ is a contant satisfying $E_0=4L_t^2(0)-\frac{4\beta_0}{L(0)^2}+\frac{C_1}{M}\frac{\alpha^2}{L^4(0)} -\frac{2C_2}{3M}\frac{\alpha^4}{L^6(0)}$. \\
Let us rewrite the right-hand side of the above equation (\ref{y2}) by substituting $y=L^2$ and $E_0$ into the equation
\begin{equation*}
\hskip-.0in
(y_t)^2=
4L_t^2(0)+4\beta_0\left(1-(\frac{L(t)}{L(0)})^2\right)-\frac{C_1}{M}\left(\frac{\alpha}{L(t)}\right)^2 \left(1-(\frac{L(t)}{L(0)})^4\right)+\frac{2C_2}{3M}\left(\frac{\alpha}{L(t)}\right)^4\left(1-(\frac{L(t)}{L(0)})^6\right).
\end{equation*}
When $L$ approaches zero, $\frac{L(t)}{L(0)} $ is very small and in the regime of $\frac{\alpha}{L(t)}\ll1$, which is required assumption for the expansion, the right-hand side will remain positive
when $L_t(0)$ is large. In other words, contrary to the equation of first order expansion (\ref{y1}), the second order expansion equation (\ref{L2}) might blow up with certain large $L_t(0)$, we will
see this is also verified in the numerical computation in the next section.

\subsection{Third order expansion} \label{third-order}
With finite time blow up in the second order expansion, one might try to
include a higher order term in the expansion of the solution
of the Helmholtz equation:
\begin{equation}
u(x)=|v(x)|^2+\alpha^2 \lp |v(x)|^2+\alpha^4 \lp^2 |v(x)|^2+\mathcal{O}( \alpha^6).
\end{equation}
As a result we have
\begin{equation*}
u_R(x)=\frac{1}{L^2}R^2(\frac{x}{L}) +(\frac{\alpha}{L})^2 \lp_x R^2(\frac{x}{L})
+(\frac{\alpha}{L})^4\lp_x^2 R^2(\frac{x}{L})+\mathcal{O} ( (\frac{\alpha}{L})^6))
\end{equation*}
and then
\begin{eqnarray*}
F(\psi_R)(x)
&=&(\lp_x u_R(x)) \psi_R(x)    \\
&\sim&  \left(\frac{1}{L^2} \lp_x R^2(\frac{x}{L}) +\frac{\alpha^2}{L^2} \lp^2_x R^2(\frac{x}{L}) +
\frac{\alpha^4}{L^4} \lp_x^3 R^2(\frac{x}{L})\right) \frac{1}{L}R(\frac{x}{L})\exp{(iS)}.
\end{eqnarray*}
Now $f_1$ satisfies the following expression
\begin{eqnarray*}
f_1
&=& \frac{L}{\pi} Re \int_{\R^2} F( \psi_R(x)) \exp {(-iS)} ( R(\rho)+\rho R_{\rho}) dx_1dx_2 \\
&\sim & I_1+I_2 +I_3,
\end{eqnarray*}
where $I_1, I_2$ are the same as in the first and second order expansion (\ref{I1}) and (\ref{I2}), and
we have
\begin{eqnarray}
I_3
&=&\frac{L}{\pi}\int_{\R^2} \frac{\alpha^4}{L^2}(\lp_x^3 R^2(\frac{x}{L}) ) \frac{1}{L}R(\frac{x}{L})
(R(\rho)+\rho R_{\rho}) dx_1dx_2     \nonumber  \\
&=&\frac{1}{\pi}\frac{\alpha^4}{L^6}\int_{\R^2} (\lp_{\xi}^3 R^2) R(R+\rho R_{\rho})\: d\xi_1\: d\xi_2  \nonumber \\
&=&-\frac{\alpha^4 C_3}{L^6},   \label{I3}
\end{eqnarray}
where the constant
\begin{equation}
\hskip-.8in
C_3=\frac{2}{\pi}\int_{\R^2} ( \gd \lp R^2)^2 d\xi_1d\xi_2>0,\quad
\xi=\frac{x}{L}=(\xi_1,\xi_2), \label{C3}
\end{equation}
where the detailed calculation is presented in claim 3 of the appendix.\\

Substituting $f_1$ into the reduced equation of $L$ (\ref{SH-redu}), we get
\begin{equation}
-L^3L_{tt}=\beta_0- \frac{C_1}{2M} \frac{\alpha^2}{L^2} + \frac{C_2}{2M} \frac{\alpha^4}{L^4}
-\frac{C_3}{2M}\frac{\alpha^6}{L^6}.    \label{L3}
\end{equation}
So $y=L^2$ satisfies the following equation
\begin{equation}
(y_t)^2=4\beta_0-\frac{\alpha^2 C_1}{M}\frac{1}{y} +\frac{2}{3}\frac{\alpha^4 C_2}{M}\frac{1}{y^2}
-\frac{\alpha^6C_3}{2M}\frac{1}{y^3}+F_0 y,   \label{y3}
\end{equation}
where $F_0=4L^2_t(0)-\frac{4\beta_0}{L^2(0)} +\frac{\alpha^2C_1}{M}\frac{1}{L^4(0)}
-\frac{2}{3}\frac{\alpha^4 C_2}{M}\frac{1}{L^6(0)} +\frac{\alpha^6C_3}{2M} \frac{1}{L^8(0)}$.
In this equation, since $C_3$ is positive, we can see again that $y$ can not approach zero, i.e., $L$ can not go to zero, which prevents singularity formation.\\
\section{Numerical Results}\label{nume}

In this section we will show
some numerical results of the evolution of $L(t)$ in three expansion cases: (\ref{L1}), (\ref{L2}) and (\ref{L3}) in order to study the prevention of blow up of the Schr\"{o}dinger-Helmholtz system.
We will first consider the first order expansion case.\\\\
{\bf{First order expansion}}\\
Let us look at the equation (\ref{L1}).
After some algebraic calculation and calculus integration, we come up with equation (\ref{y1}):
\begin{equation*}
\hskip-.8in
(y_t)^2 =4\beta_0-\frac{\alpha^2C_1}{M} \frac{1}{y}+4D_0y,
\end{equation*}
where $\beta_0=\beta(0)+\frac{ C_1}{2M}\frac{\alpha^2}{L^2(0)}$ and  $D_0=L_t^2(0)-\frac{\beta_0}{L^2(0)}+\frac{C_1}{4M} \frac{\alpha^2}{L^4(0)}$. \\

>From above we know that $y$ can not approach zero, equivalently, $L$ can not
go to zero, which prevents singularity formation. \\

Furthermore, Fibich and Papanicolaou \cite{FP} derived the generic equation
\begin{equation*}
\hskip-.8in
(y_t)^2 =4\beta_0-\frac{\alpha^2C_1}{M} \frac{1}{y}+4\frac{H_0}{M}y=\frac{-4H_0}{M}\frac{1}{y}(y_M-y)(y-ym),
\end{equation*}
where
\begin{eqnarray*}
\hskip-1in
&& y_M=\frac{ \sqrt { \beta_0^2+\alpha^2C_1H_0/M^2}+\beta_0 }{-2H_0/M}
=\frac{M\beta_0}{-H_0}
[1+\mathcal{O} (\frac{\alpha^2H_0}{\beta_0^2} )], \\
&&y_m=\frac{\alpha^2C_1}{2M}\frac{1}{ \sqrt { \beta_0^2+\alpha^2C_1H_0/M^2}+\beta_0} =
\frac{\alpha^2C_1}{4M\beta_0} [1+\mathcal{O }(\frac{\alpha^2 H_0}{\beta_0^2} )] ,\\
&&H_0=H(0)+\frac{\alpha^2C_1}{4}\frac{1}{L^4(0)}.
\end{eqnarray*}
>From above we see that when $H_0>0$, and $L_t(0)>0$, $L$ is monotonically defocusing to infinity;
when $H_0>0$ and $L_t(0)<0$, self-focusing is arrested when $L=L_m=(y_m)^{1/2}>0$, after which
$L$ is monotonically defocusing to infinity; when $H_0<0$, then $L$ goes through periodic oscillation between $L_m=(y_m)^{1/2}$ and $L_M=(y_M)^{1/2}$ (see Figure 1).\\\\

\noindent
{\bf{Second order expansion}}\\
Now we look at the equation (\ref{L2}) of next order expansion:
\begin{equation}
\hskip-.8in
-L^3L_{tt}=\left(\frac{C_2}{2M}\right)\left(\frac{\alpha}{L}\right)^4-\left(\frac{C_1}{2M}\right) \label{L22}
\left(\frac{\alpha}{L}\right)^2+\beta_0.
\end{equation}
In this equation, when $\beta_0>\frac{C_1^2}{C_2}\frac{1}{8M}$, the right-hand side is definitely
positive, then for any initial data $L(0)$ with $L_t(0)<0$, the solution $L$ will monotonically decrease
and approach zero in finite time. In other words, when the initial excess energy is larger than certain
amount ($\frac{C_1^2}{C_2}\frac{1}{8M}$) and $L$ is initially focusing, $L$ will focus and blow up
in finite time (see Figure 2). However, this is not valid to begin with applying the modulation theory.
Recall that for us to apply the modulation theory to a perturbed critical NLS, we require three conditions
to hold. One of the conditions is to require $|\beta(t)|\ll1$ (see \ref{condition2} or \ref{condition22}).\\

So we will consider the case when $0<\beta_0<\frac{C_1^2}{C_2}\frac{1}{8M}$, then we have $\:r_{\mbox{low}}=\sqrt{\frac{ C_1-2M\sqrt{K}}{2C_2}},\:
r_{\mbox{high}}=\sqrt{   \frac{C_1+2M\sqrt{K}   }{2C_2}    }$, where $K=\frac{C_1^2}{C_2}\frac{1}{8M}-\beta_0$. In this case, when initially $\frac{\alpha}{L(0)}>r_{\mbox{high}}$ and $L_t(0)<0$, the right-hand side of the equation (\ref{L22}) will remain positive, so
$L$ will monotonically decrease to zero, which is similar to the case of $\beta_0>\frac{C_1^2}{C_2}\frac{1}{8M}$. Once again, this is not valid here for the discussion since
the asymptotic expansion (\ref{second-order}) is valid under
the assumption that $\frac{\alpha}{L} \ll1$, so we need only to consider the situation of
$\frac{\alpha}{L(0)}$ small, in this case, $\frac{\alpha}{L(0)}<r_{\mbox{high}}$.\\\\

Finally, we consider only the case $0<\beta_0<\frac{C_1^2}{C_2}\frac{1}{8M}$ and $0<\frac{\alpha}{L(0)}<r_{\mbox{high}}$. \\

First we focus on $0<\frac{\alpha}{L(0)}<r_{low}$. In this case,
$L$ might defocus to infinity, oscillate between two values or even blow up in finite time depending on different initial condition of $L_t(0)$ for given
$\beta_0,\alpha$ and $L(0)$. In Figure 3, we take the parameters $\alpha=0.01, \: \frac{\alpha}{L(0)}=r_{low}/2$ and $ \beta_0=0.01$. When initially $L_t(0)>-44.9999$, $L$ will eventually defocus to infinity
if $H_0>0$ (3a) and $L$ will oscillate between two values if $H_0<0$ (3b); when initially $L_t(0)<-49.9999$, $L$ will approach zero in finite time, i.e., we observe singularity in finite time (3c). This numerical result is expected from analysis at the end of subsection (\ref{next-order}).\\

Similarly, for $r_{low}<\frac{\alpha}{L(0)}<r_{high}$, we will observe different behaviors - defocusing, oscillation or focusing depending on $L_t(0)$ and $H_0$. For instance, when $\frac{\alpha}{L(0)}=\frac{r_{low}+r_{high}}{2}$, we have the threshold value $L_t^c=- 39.9999$, i.e., when $L_t(0)>-39.9999$,
$L$ will eventually defocus to infinity if $H_0>0$ and oscillate between two values if $H_0<0$; when
$L_t(0)<-39.9999$, $L$ will eventually decrease to zero.\\\\

\noindent
{\bf{Third order expansion}}\\
Lastly we study the equation (\ref{L3}) of higher order expansion:
\begin{equation*}
-L^3L_{tt}=\beta_0- \frac{C_1}{2M} \frac{\alpha^2}{L^2} + \frac{C_2}{2M} \frac{\alpha^4}{L^4}
-\frac{C_3}{2M}\frac{\alpha^6}{L^6}.
\end{equation*}
By defining $y=L^2$ and integrating the equation, we get equation (\ref{y3})
\begin{equation*}
(y_t)^2=4\beta_0-\frac{\alpha^2 C_1}{M}\frac{1}{y} +\frac{2}{3}\frac{\alpha^4 C_2}{M}\frac{1}{y^2}
-\frac{\alpha^6C_3}{2M}\frac{1}{y^3}+F_0 y,
\end{equation*}
where $F_0=4L^2_t(0)-\frac{4\beta_0}{L^2(0)} +\frac{\alpha^2C_1}{M}\frac{1}{L^4(0)}
-\frac{2}{3}\frac{\alpha^4 C_2}{M}\frac{1}{L^6(0)} +\frac{\alpha^6C_3}{2M} \frac{1}{L^8(0)}$.\\
In the above equation (\ref{y3}), $y$ can not approach zero since the leading order
term on the right-hand side is of negative sign when $y$ goes to zero, equivalently, $L$ can not
approach zero. The nature of this equation is the same as that of equation (\ref{y1}), and we
see the same pattern in the numerical result (see Figure 4).\\\\
\section{conclusion}\label{conclusion}
>From the analysis and numerical computation, we see that the regularization of the classical NLS
effectively prevents singularity formation with positive parameter $\alpha>0$. By asymptotically expanding the solution of the Helmholtz equation to approximate the reduced system of the
modulation theory, we observe strong no blow-up pattern in both first order and third order
expansion. In the valid regime of the expansion and modulation theory, we also observe
no blow-up pattern in the second order expansion with further restriction on certain condition:
$L_t(0)>L_t^c$, threshold initial value of $L_t(0)$. This phenomenon is expected for even
higher order expansion, say fourth order expansion approximation. One of the reasons is that
the Laplace operator is not bounded, which causes instability for the expansion of the solution
of the Helmholtz equation.

\begin{figure}[htp]
\centering
\includegraphics[width=3in,height=2.5in]{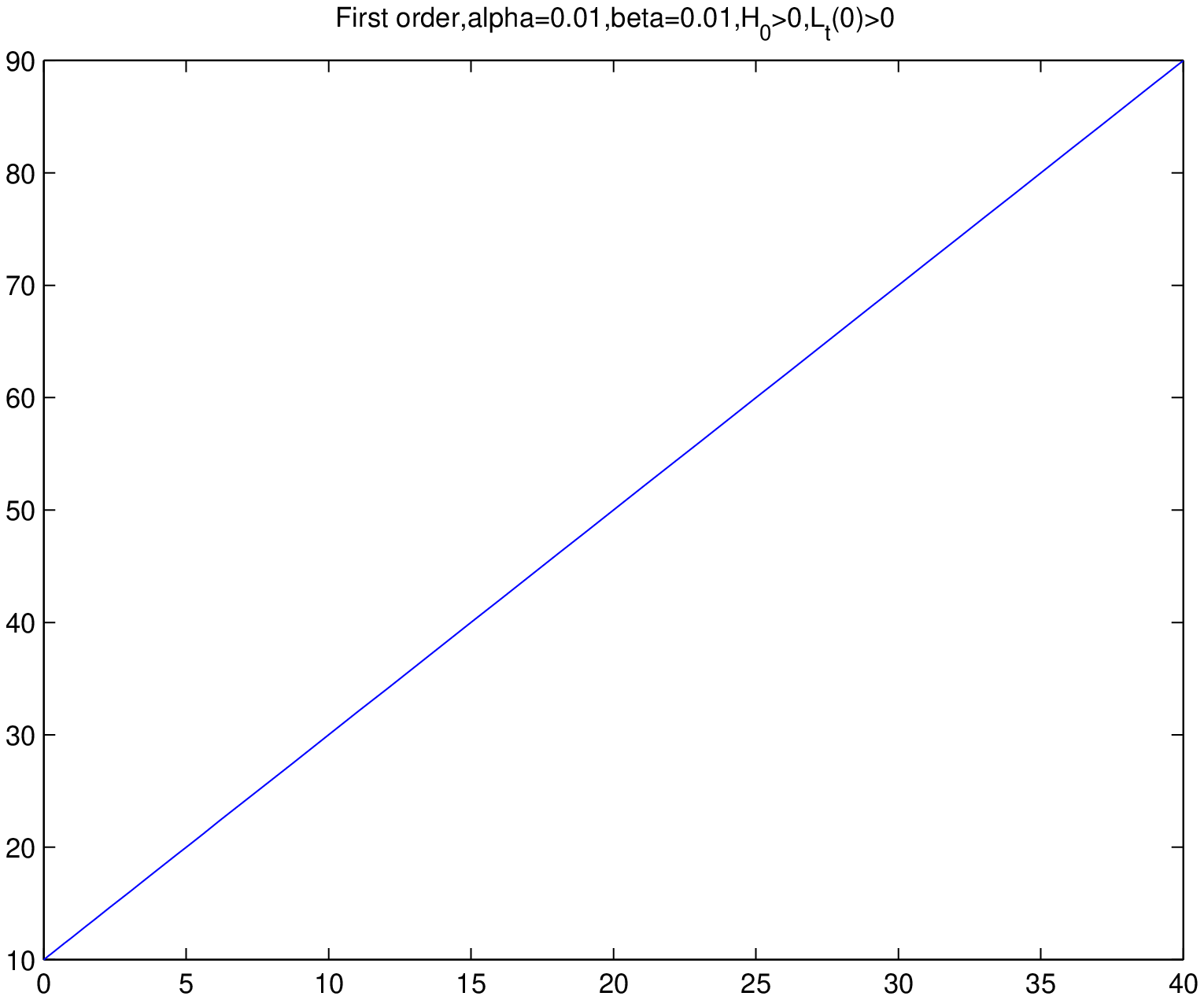}$1a$
\includegraphics[width=3in,height=2.5in]{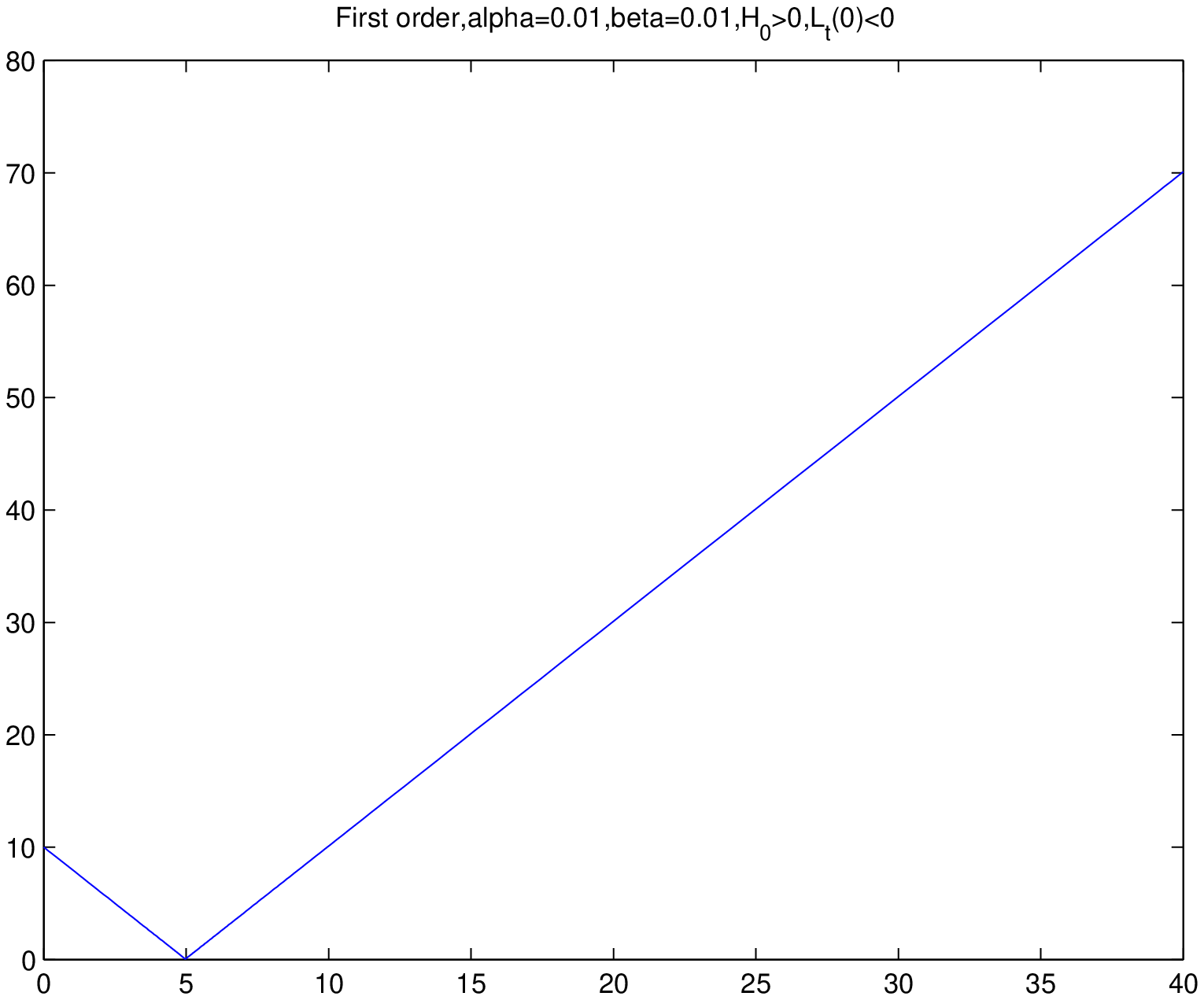}$1b$
\includegraphics[width=3in,height=2.5in]{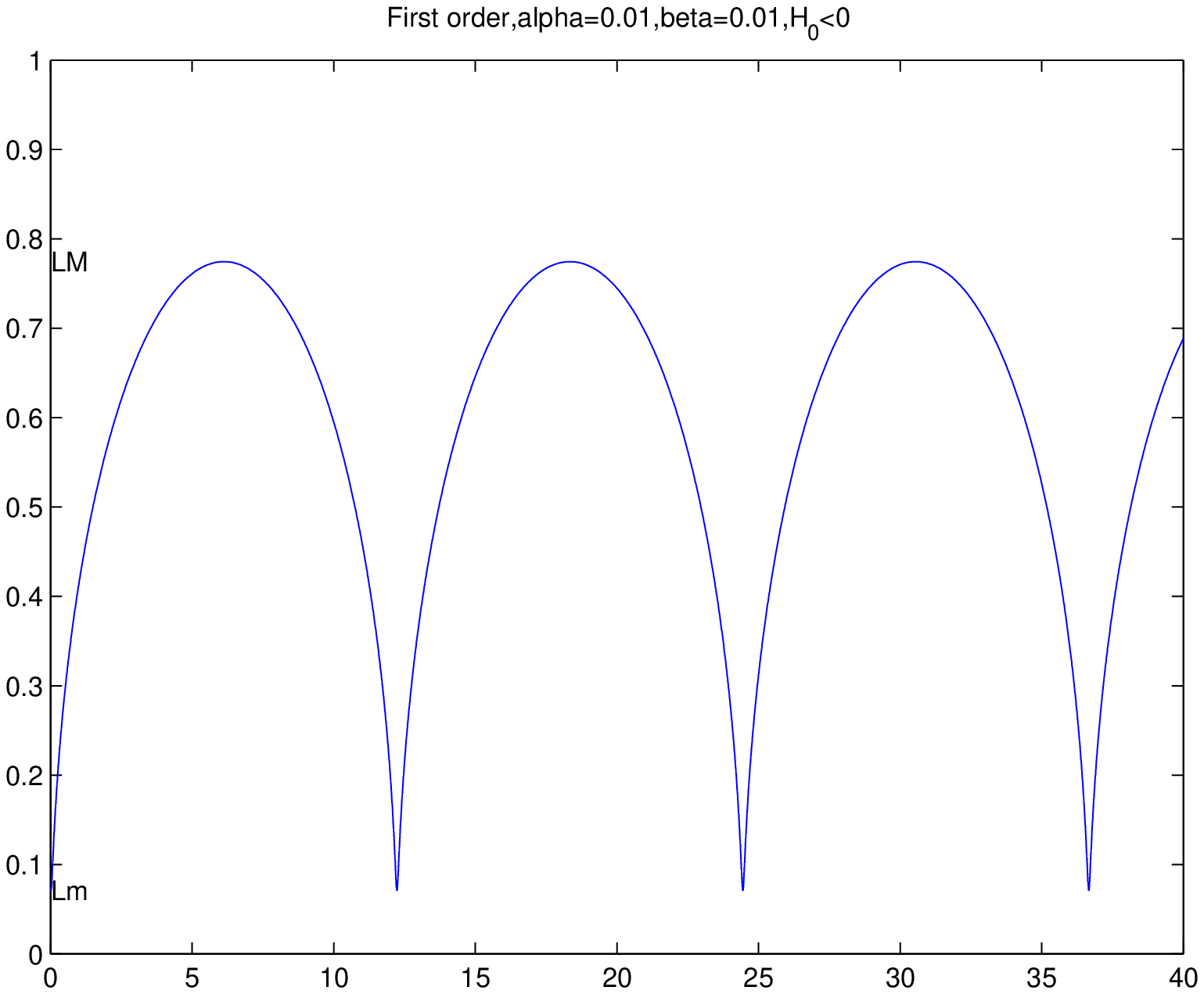}$1c$
\caption{$L$ evolves in time in first order expansion. $1a$ monotonic defocusing, $H_0>0,L_t(0)>0$. $1b$  first focusing then defocusing, $H_0>0,L_t(0)<0$.  $1c$ oscillation, $H_0<0$. For all cases, $\alpha=0.01,\beta_0=0.01$. $1a,1b$, $\frac{\alpha}{L(0)}=0.001$. In $1c$, $\frac{\alpha}{L(0)}=1/8$.}
\label{fig:firstorderL}
\end{figure}

\begin{figure}[htp]
\centering
\includegraphics[width=3in,height=2.5in]{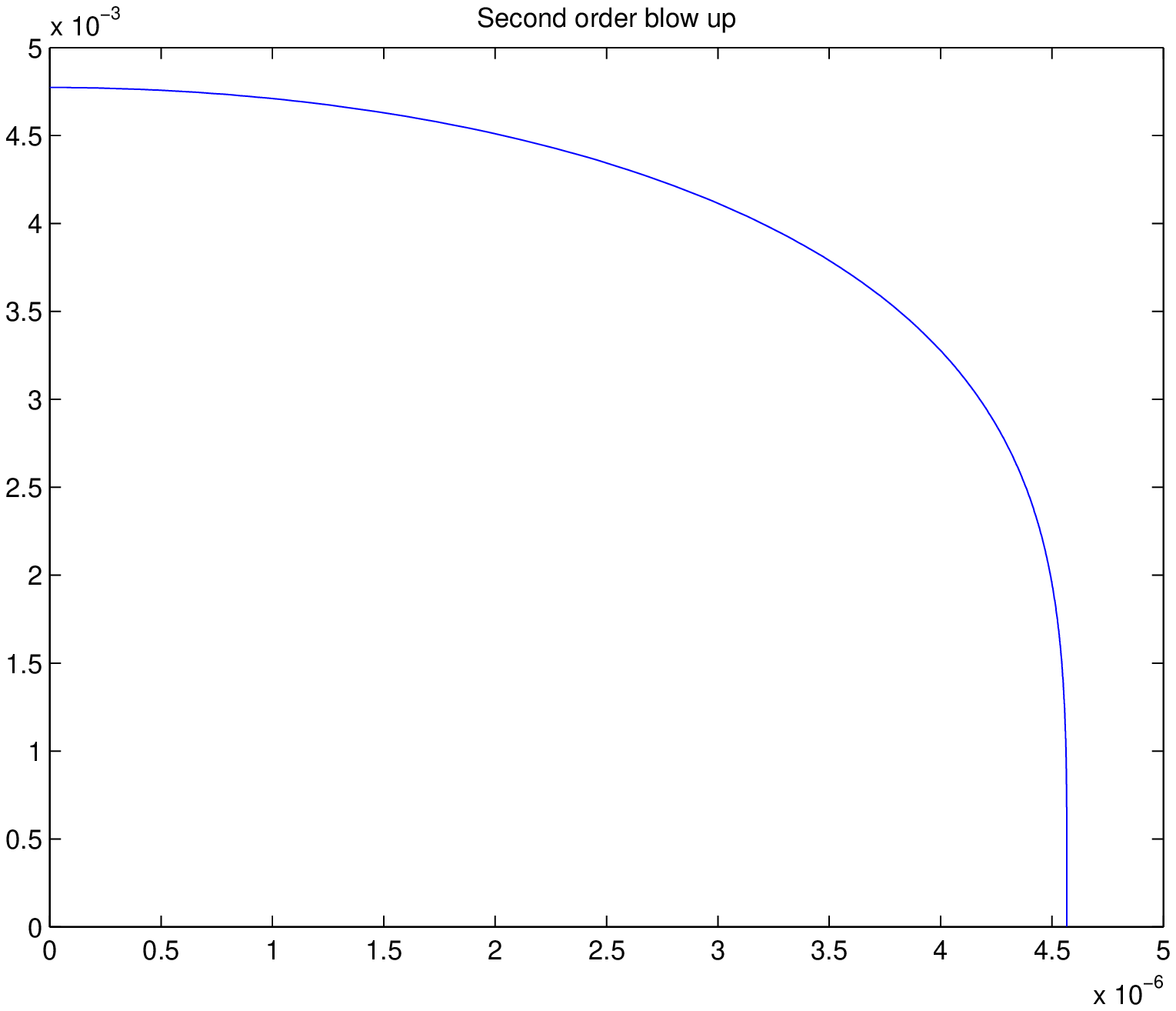}$2$
\caption{Flow of $L$ in second order expansion. $L$ decreases to zero in finite tim, $\alpha=0.01,\beta_0>>1,L(0)=0.1,L_t(0)=-2$}
\label{fig:secondorderL}
\end{figure}

\pagebreak

\begin{figure}[htp]
\centering
\includegraphics[width=3in, height=2.5in]{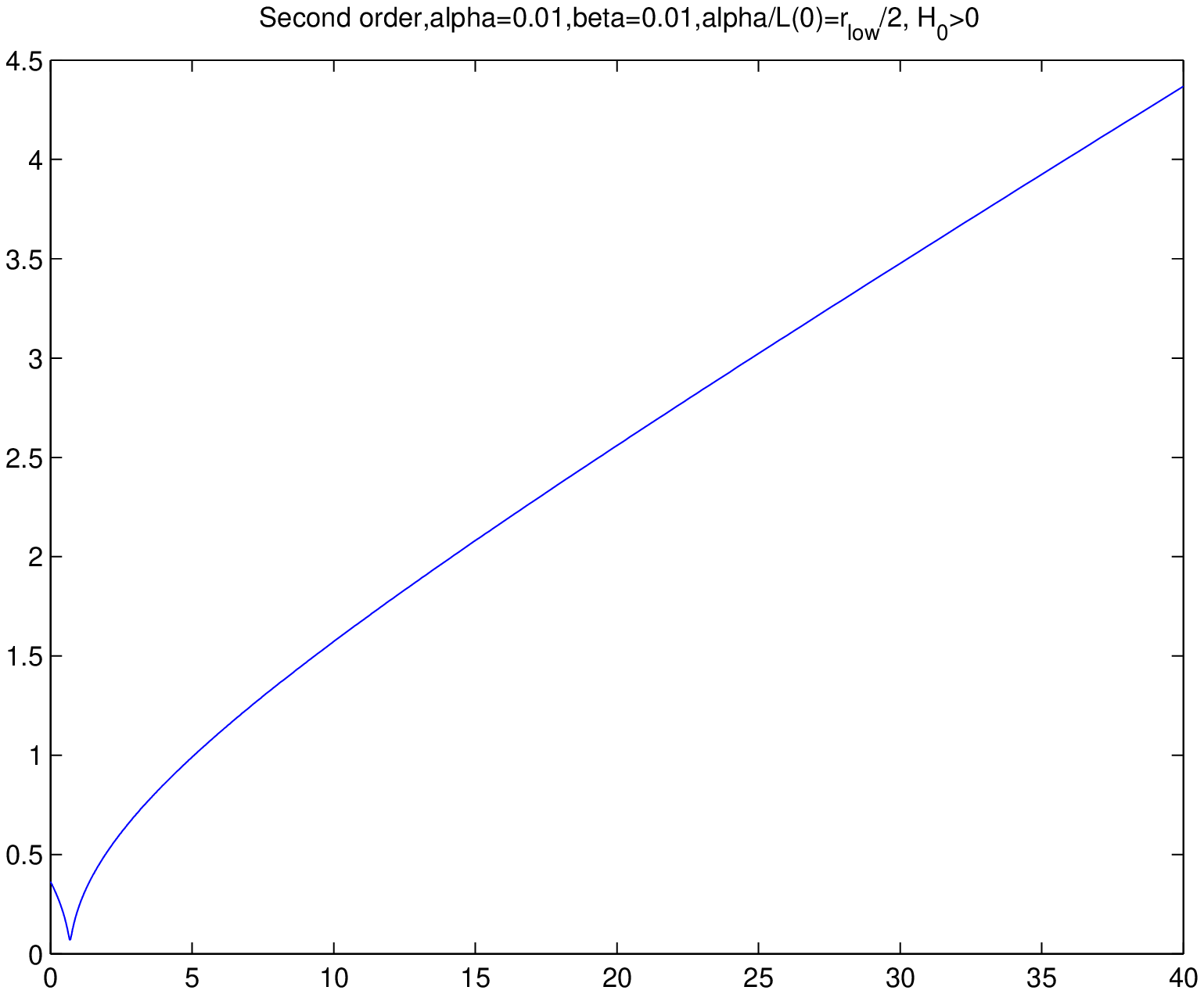}$3a$
\includegraphics[width=3in,height=2.5in]{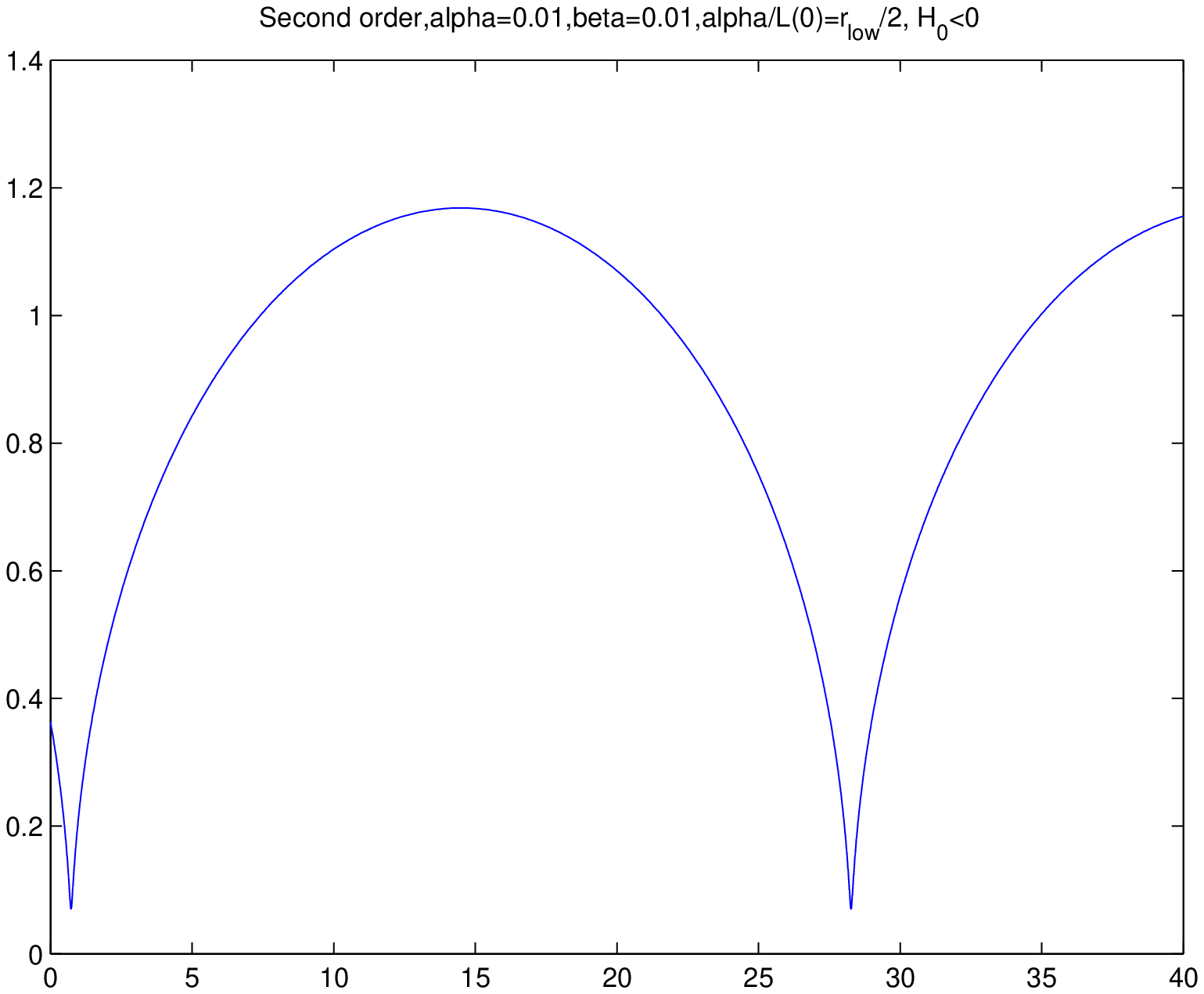}$3b$
\includegraphics[width=3in, height=2.5in]{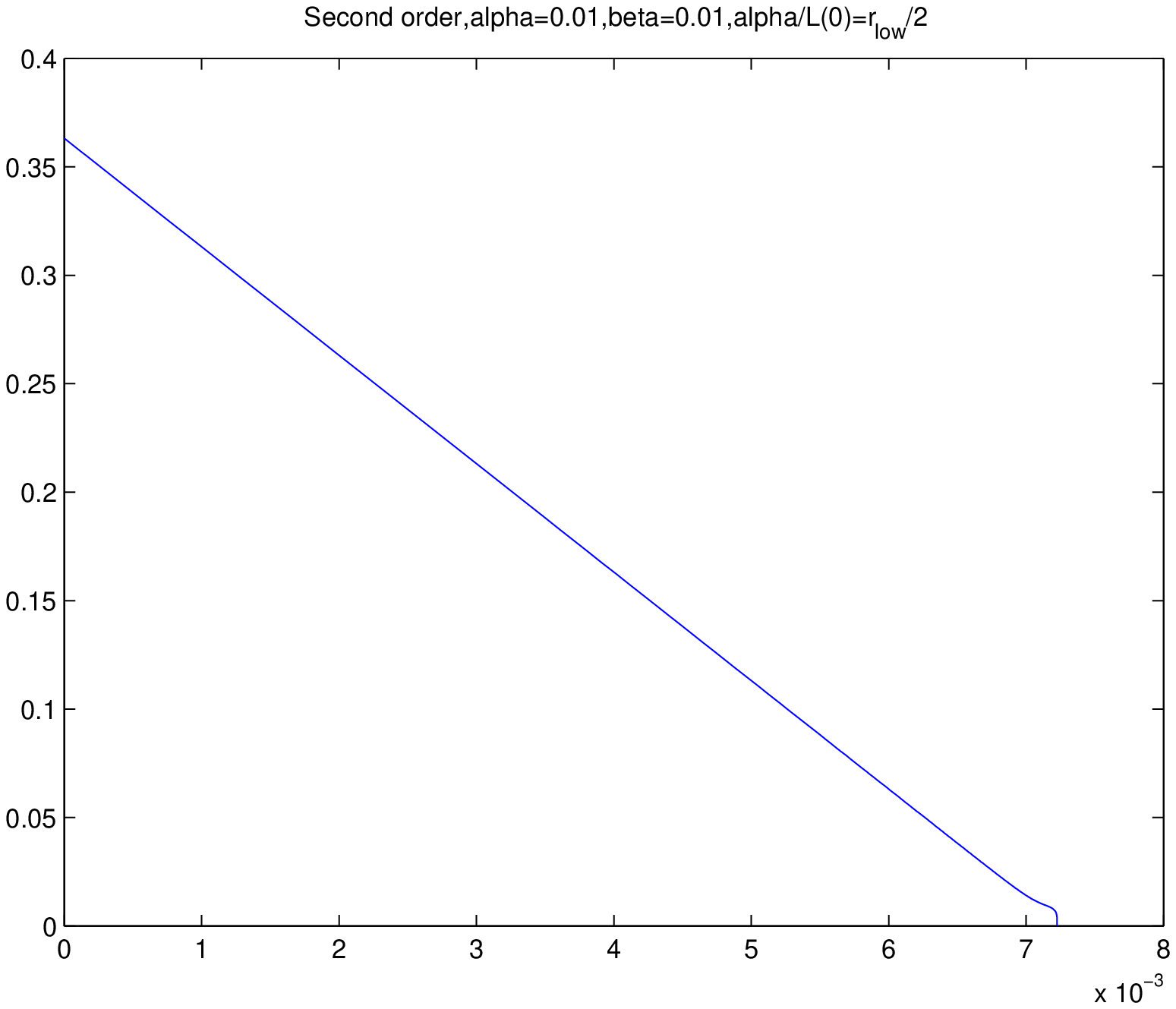}$3c$
\caption{Flow of $L$ in second order expansion. $3a$ $L$ defocuses to infinity when $H_0>0$. $3b$ $L$ oscillates between two values
when $H_0<0$. $3c$ $L$ blows up when $|L_t(0)|$ is large. For all cases, we use $\alpha=0.01, \beta_0=0.01<<\frac{C_1^2}{C_2}\frac{1}{8M}, \alpha/L(0)=r_{low}/2$.}
\label{fig:secondorder}
\end{figure}

\begin{figure}[htp]
\centering
\includegraphics[width=3in,height=2.5in]{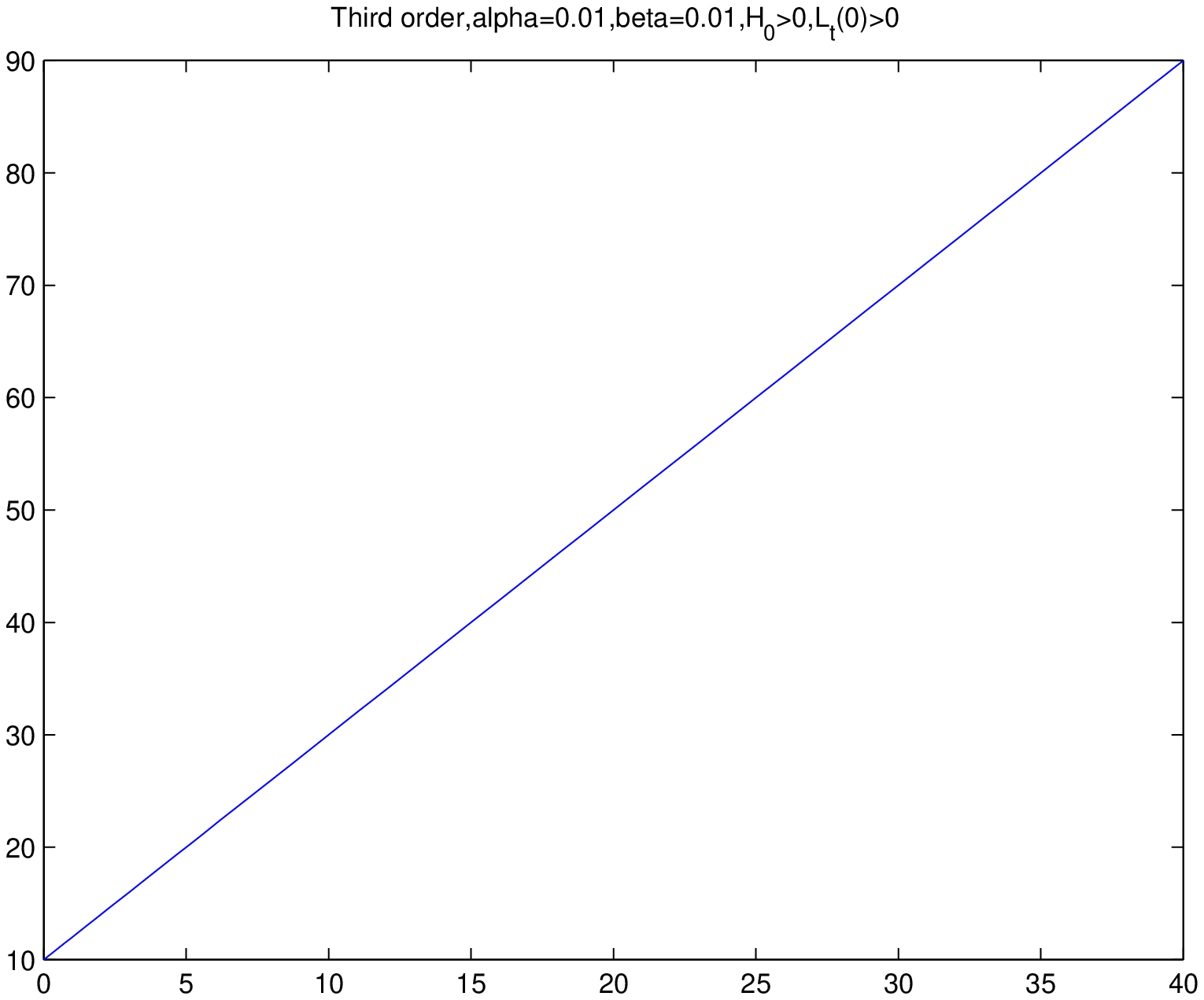}$4a$
\includegraphics[width=3in,height=2.5in]{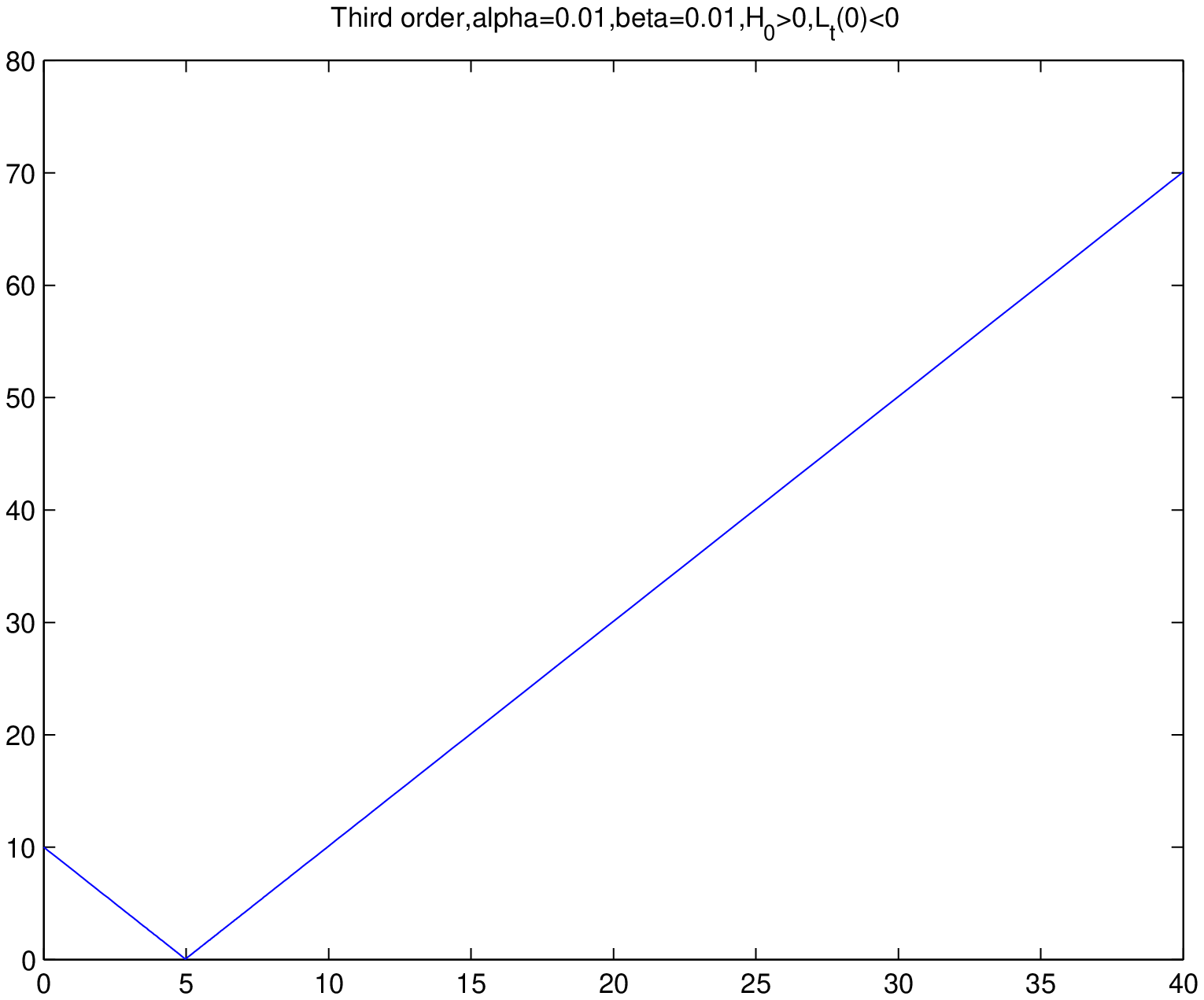}$4b$
\includegraphics[width=3in,height=2.5in]{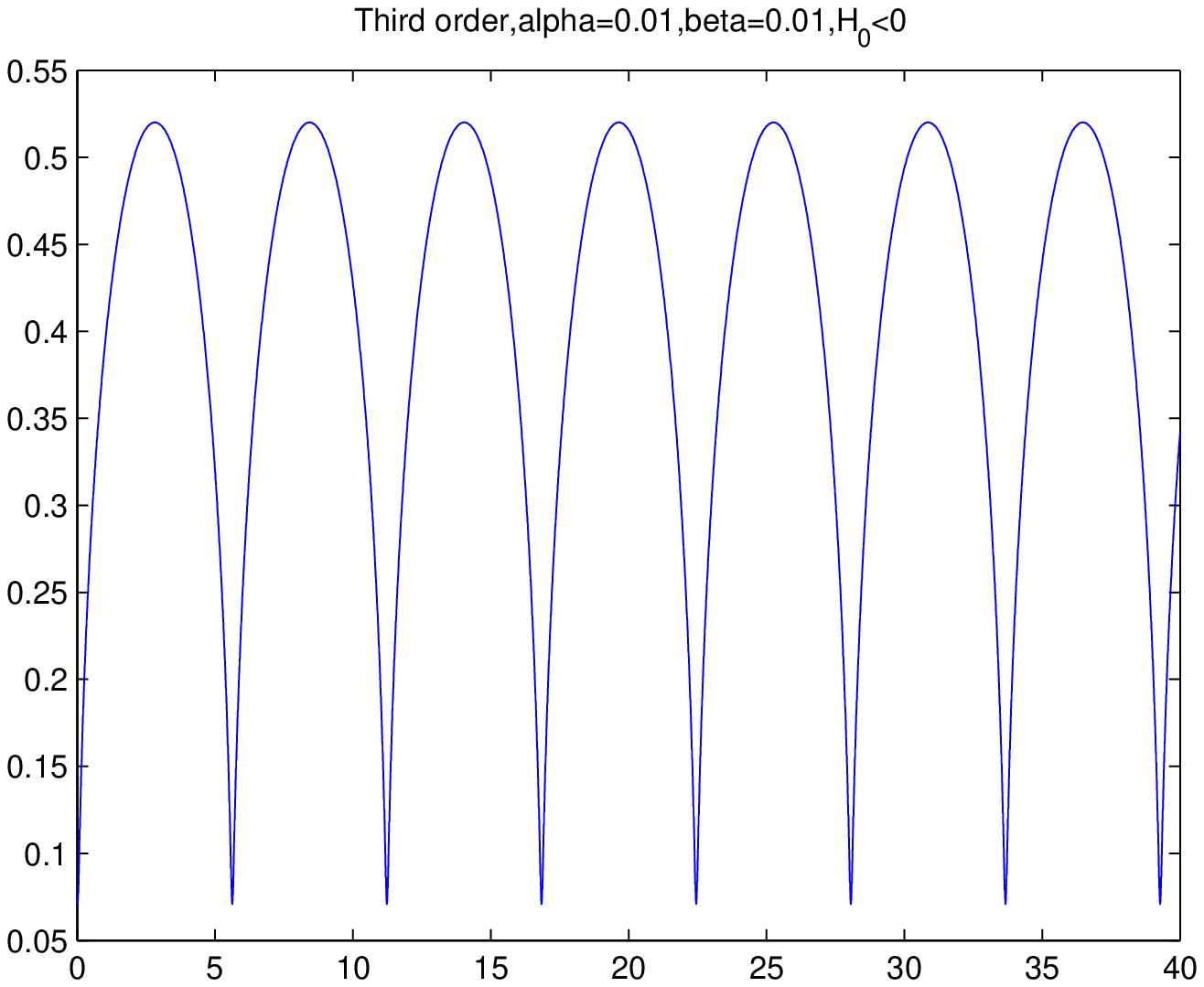}$4c$
\caption{Flow of $L$ in third order expansion. $4a$ monotonic defocusing, $H_0>0,\:L_t(0)>0$. $4b$ first focusing then defocusing, $H_0>0,\:L_t(0)<0$. $4c$ oscillation between two values, $H_0<0$. For all cases, $\alpha=0.01,\beta_0=0.01$
In $4a$ and $4b$, $\frac{\alpha}{L(0)}=0.001$. In $4c$, $\frac{\alpha}{L(0)}=1/8$.}
\label{fig:thirdorderL}
\end{figure}

\pagebreak

\pagebreak

\section*{appendix}
For completeness, we present in this section the detail of the calculation of the
integrals (\ref{I1}), (\ref{I2}) and (\ref{I3}).\\

{\bf{Claim 1}}:
The integral $I_1$ in (\ref{I1}) for the first order expansion can be simplified as
\begin{equation*}
\hskip-.8in
I_1=\frac{1}{\pi}\frac{1}{L^2}\int_{\R^2}(\lp_{\xi}R^2)R(R+\rho R_{\rho})\:d\xi_1d\xi_2
=-\frac{2}{L^2}\int_0^\infty [ (R^2)_{\rho}]^2\rho \:d\rho
\end{equation*}
\begin{proof}
\begin{eqnarray*}
\hskip-.8in
I_1
&=&\frac{1}{\pi}\frac{1}{L^2} \int_{\R^2}(\lp_{\xi}R^2(\rho))R(\rho)(R(\rho)+\rho R_{\rho})\:d\xi_1\:d\xi_2 \\
&=&\frac{1}{\pi}\frac{1}{L^2} \int_{\R^2}(\lp R^2) R^2 \:d\xi_1 \: d\xi_2
+ \frac{1}{\pi}\frac{1}{L^2}\int_{\R^2}(\lp R^2)R\: R_{\rho} \:\rho \:d\xi_1\: d\xi_2  \\
&=&I_{11}+I_{12}.
\end{eqnarray*}
In the rest of this section, all the integrands and integrals are of variables $\xi$ or $\rho$ (i.e., they are the scaled variables) unless it is stated otherwise.\\

Now for $I_{11}$, we integrate by parts once and change the variables by $\xi_1=\rho cos\theta,\:
\xi_2=\rho sin\theta$, we obtain
\begin{eqnarray*}
\hskip-.8in
I_{11}
&=&-\frac{1}{\pi}\frac{1}{L^2}\int_{\R^2} (\gd R^2)^2 \: d\xi_1 \: d\xi_2 \\
&=& -\frac{1}{\pi}\frac{1}{L^2} 2\pi \int_0^{\infty}[ (R^2)_{\rho}]^2 \: \rho \: d \rho \\
&=& -\frac{2}{L^2} \int_0^{\infty} [ (R^2)_{\rho}]^2 \: \rho \: d\rho,
\end{eqnarray*}
which gives us exactly the constant $C_1$ as in  (\ref{C1}).\\

For $I_{12}$, we show now it is identically zero.\\
Using polar coordinates, we get
\begin{eqnarray*}
\hskip-.8in
I_{12}
&=&\frac{1}{\pi}\frac{1}{L^2} 2\pi \int_0^{\infty} ( (R^2)_{\rho\rho}+\frac{1}{\rho}(R^2)_{\rho})R
R_{\rho}\: \rho^2 \: d\rho   \\
&=&\frac{2}{L^2} \int_0^{\infty}(R^2)_{\rho\rho} R R_{\rho}\: \rho^2 \: d\rho+\frac{2}{L^2}\int_0^{\infty}
(R^2)_{\rho}RR_{\rho} \rho \: d\rho   \\
&=& \frac{1}{L^2}\int_0^{\infty}(R^2)_{\rho\rho}(R^2)_{\rho} \:\rho^2 \: d\rho+\frac{1}{L^2}\int_0^{\infty}
(R^2)_{\rho}(R^2)_{\rho}\: \rho \: d\rho  \\
&=&X+Y.
\end{eqnarray*}
For the first integral $X$, after rewriting and integration by parts once, we get
\begin{eqnarray*}
\hskip-.8in
X
&=&\frac{1}{L^2}\int_0^{\infty}\frac{1}{2}[ ((R^2)_{\rho})^2]_{\rho} \:\rho^2 d\rho \\
&=& -\frac{1}{L^2}\int_0^{\infty}((R^2)_{\rho})^2\: \rho \: d\rho \\
&=& -Y.
\end{eqnarray*}
We conclude that $I_{12}=0$, which yields the result of (\ref{I1}) and (\ref{C1}).
\end{proof}

{\bf{Claim 2}}
The integral $I_2$ (\ref{I2}) in the next order expansion can be simplified as
\begin{equation*}
\hskip-.8in
I_2=\frac{1}{\pi}\frac{\alpha^2}{L^4}\int_{\R^2}(\lp^2_{\xi}(R^2))R(R+\rho R_{\rho}) \:d\xi_1 d\xi_2
=\frac{3\alpha^2}{2\pi L^4}\int_{\R^2}(\lp_{\xi}R^2)^2\:d\xi_1d\xi_2
\end{equation*}
\begin{proof}
\begin{eqnarray*}
\hskip-.8in
I_2
&=&\frac{1}{\pi}\frac{\alpha^2}{L^4}\int_{\R^2} (\lp^2(R^2))R(R+\rho R_{\rho}) \: d\xi_1\: d\xi_2 \\
&=&\frac{1}{\pi}\frac{\alpha^2}{L^4}\int_{\R^2}(\lp^2(R^2))R^2\: d\xi_1\: d\xi_2 +
\frac{1}{\pi}\frac{\alpha^2}{L^4}\int_{\R^2}(\lp^2(R^2)) R( \rho R_{\rho}) \: d\xi_1\: d\xi_2 \\
&=&I_{21}+I_{22}.
\end{eqnarray*}
After integration by parts twice, we get
\begin{equation*}
\hskip-.8in
I_{21}=\frac{1}{\pi}\frac{\alpha^2}{L^4}\int_{\R^2}[ \lp(R^2)]^2 \: d\xi_1 \: d\xi_2.
\end{equation*}
Next we will show that $I_{22}= \frac{1}{2}I_{21}$, which yields (\ref{C2}).\\

Recall that $\rho R_{\rho}=\xi \cdot \gd R,\:\xi=(\xi_1,\xi_2),\:\rho^2=\xi_1^2+\xi_2^2$, so we have
\begin{eqnarray*}
\hskip-.8in
I_{22}
&=&\frac{1}{\pi}\frac{\alpha^2}{L^4}\int_{\R^2}( \lp^2(R^2))(R (\xi \cdot \gd R)) \: d\xi \\
&=&\frac{1}{\pi}\frac{\alpha^2}{L^4}\int_{\R^2}( \lp^2(R^2))(\xi \cdot \gd \frac{1}{2}R^2)\: d\xi \\
&=&\frac{1}{\pi}\frac{\alpha^2}{L^4}\int_{\R^2} \lp(R^2)\lp (\xi \cdot \gd \frac{1}{2}R^2) \: d\xi \\
&=& \frac{1}{\pi}\frac{\alpha^2}{L^4}\int_{\R^2} \lp(R^2)\left(\lp (R^2)+\xi \cdot \gd(\lp \frac{1}{2}R^2)\right)\: d\xi \\
&=&\frac{1}{\pi}\frac{\alpha^2}{L^4}\int_{\R^2} [\lp(R^2)]^2\: d\xi+\frac{1}{\pi}\frac{\alpha^2}{L^4}\int_{\R^2} \lp(R^2)(\xi \cdot \gd (\lp \frac{1}{2}R^2) )\: d\xi \\
&=&P+Q.
\end{eqnarray*}
For the second integral in the last line, we rewrite it then integrate by parts and obtain
\begin{eqnarray*}
\hskip-1in
Q
&=&\frac{1}{\pi}\frac{\alpha^2}{L^4}\int_{\R^2} \lp(R^2)(\xi \cdot \gd ( \lp \frac{1}{2}R^2) )\: d\xi  \\
&=& \frac{1}{\pi}\frac{\alpha^2}{L^4}\int_{\R^2} \xi \cdot \gd  \left(\frac{1}{4} (\lp(R^2))^2\right) \: d\xi  \\
&=&-\frac{1}{\pi}\frac{\alpha^2}{L^4}\int_{\R^2} (\gd \cdot \xi )\frac{1}{4} (\lp(R^2))^2 \: d\xi \\
&=& -\frac{1}{\pi}\frac{\alpha^2}{L^4}\int_{\R^2}\frac{1}{2} [\lp(R^2)]^2\: d\xi \\
&=&-\frac{1}{2}P.
\end{eqnarray*}
So we get $I_{22}=P+Q=\frac{1}{2}P=\frac{1}{2}I_{21}$, so we conclude
that $I_2=\frac{3}{2}I_{21}=\frac{\alpha^2}{L^4}\frac{3}{2\pi}\int_{\R^2}[ \lp(R^2)]^2\: d\xi$, which yields
exactly (\ref{I2}) and (\ref{C2}).
\end{proof}

{\bf{Claim 3}}
The integral $I_3$ (\ref{I3}) in the calculation of higher order expansion can be simplified as
\begin{equation*}
\hskip-.8in
I_3=\frac{1}{\pi}\frac{\alpha^4}{L^6}\int_{\R^2}(\lp^3_{\xi}R^2)R(R+\rho R_{\rho} )\:d\xi_1d\xi_2
=-\frac{2}{\pi}\frac{\alpha^4}{L^6}\int_{\R^2}(\gd \lp R^2)^2\: d\xi_1d\xi_2.
\end{equation*}
\begin{proof}
\begin{eqnarray*}
\hskip-.8in
I_3
&=&\frac{1}{\pi}\frac{\alpha^4}{L^6}\int_{\R^2}(\lp^3(R^2))R(R+\rho R_{\rho}) \: d\xi \\
&=&\frac{1}{\pi}\frac{\alpha^4}{L^6}\int_{\R^2} (\lp^3(R^2))R^2 \: d\xi
+\frac{1}{\pi}\frac{\alpha^4}{L^6}\int_{\R^2} (\lp^3(R^2)) R(\rho R_{\rho})\: d\xi \\
&=& I_{31}+I_{32}.
\end{eqnarray*}
After integration by parts three times, we get
\begin{equation*}
\hskip-.8in
I_{31}= -\frac{1}{\pi}\frac{\alpha^4}{L^6} \int_{\R^2} (\gd \lp (R^2) )^2\: d\xi.
\end{equation*}
Next, we will show that $I_{32}=I_{31}= -\frac{1}{\pi}\frac{\alpha^4}{L^6} \int_{\R^2} ( \gd \lp (R^2) )^2\: d\xi$.
\begin{eqnarray*}
\hskip-.8in
I_{32}
&=& \frac{1}{\pi}\frac{\alpha^4}{L^6} \int_{\R^2} ( \lp^3(R^2))R(\rho R_{\rho})\: d\xi \\
&=&   \frac{1}{\pi}\frac{\alpha^4}{L^6} \int_{\R^2} ( \lp^3(R^2))R( \xi \cdot \gd R)\: d\xi  \\
&=& \frac{1}{\pi}\frac{\alpha^4}{L^6} \int_{\R^2} ( \lp^3(R^2)) (\xi \cdot \gd \frac{1}{2}R^2)\: d\xi \\
&=& \frac{1}{\pi}\frac{\alpha^4}{L^6}\int_{\R^2} (\lp^2(R^2)) \lp ( \xi \cdot \gd \frac{1}{2}R^2)\: d\xi \\
&=& \frac{1}{\pi}\frac{\alpha^4}{L^6}\int_{\R^2} (\lp^2(R^2)) \left(\lp (R^2)+\xi \cdot \gd(\lp \frac{1}{2}R^2)\right) \: d\xi   \\
&=& \frac{1}{\pi}\frac{\alpha^4}{L^6}\int_{\R^2} (\lp^2(R^2)) \lp(R^2) \: d\xi
+\frac{1}{\pi}\frac{\alpha^4}{L^6}\int_{\R^2}\frac{1}{2} (\lp^2(R^2)) (\xi \cdot \gd (\lp R^2))\: d\xi \\
&=& A+B.
\end{eqnarray*}
For $A$, after integration by parts once, we obtain
\begin{equation*}
\hskip-.8in
A= -\frac{1}{\pi}\frac{\alpha^4}{L^6}\int_{\R^2} (\gd \lp (R^2) )^2 \: d\xi.
\end{equation*}
For $B$, we define $\phi=\lp(R^2)$, a scalar function. Then we rewrite and calculate the term $B$ as follows
\begin{eqnarray*}
\hskip-.8in
B
&=&\frac{1}{2\pi}\frac{\alpha^4}{L^6}\int_{\R^2} \lp \phi (\xi \cdot \gd \phi )\: d\xi   \\
&=&\frac{1}{2\pi}\frac{\alpha^4}{L^6}2\pi\int_0^\infty (\phi_{\rho\rho}+\frac{1}{\rho}\phi_{\rho})( \rho\:\phi_{\rho}) \rho \: d\rho \\
&=& \frac{\alpha^4}{L^6}\int_0^\infty \phi_{\rho\rho} \phi_\rho \: \rho^2+ (\phi_{\rho})^2 \rho \: d\rho \\
&=& \frac{\alpha^4}{L^6}\int_0^\infty \frac{1}{2} ( (\phi_{\rho})^2 \rho^2 )_{\rho} \: d\rho \\
&=&0.
\end{eqnarray*}
So we have now $I_3=I_{31}+I_{32}=2I_{31}=- \frac{2}{\pi}\frac{\alpha^4}{L^6} \int_{\R^2} (\gd \lp (R^2) )^2\: d\xi$, which concludes our result of (\ref{I3}) and (\ref{C3}).
\end{proof}
\section*{Acknowledgements}
We would like to thank Professor Gadi Fibich for the valuable comments and suggestion.
This work was supported in part by the NSF grants no. DMS-0504619 and no. DMS-0708832
and the ISF grant no. 120/06.



\begin{thebibliography}{99}


\bibitem{AALRT} A. Aceves, C. De Angelis, G. Luther, A. Rubenchik and S. Turitsyn,
                {\em All-optical-switching and pulse amplification and steering in nonlinear fiber arrays},
                Physica D {\bf{87}} (1995), 262-272.



\bibitem{AART} A. Aceves, C. De Angelis, A. Rubechik and S. Turitsyn,
                       {\em Multidimensional solitons in fiber arrays}, Opt. Lett. {\bf{19}} (1994), 329-331.

\bibitem {AALRT2} A. Aceves, C. De Angelis, G. Luther, A. Rubenchik and S. Turitsyn,
            {\em Energy localization in nonlinear fiber arrays: Collapse effect compressor},
                Phys. Rev. Lett. {\bf{75}} (1995), 73-76.

\bibitem{AALRT3} A. Aceves, C. De Angelis, G. Luther, A. Rubenchik and S. Turitsyn,
           {\em Optical pulse compression using fiber arrays}, Optical Fiber Technology {\bf{1}} (1995), 244-246.


\bibitem{CMT} Y. Cao, Z. H. Musslimani and E. S. Titi, {\em Nonlinear Schr\"{o}dinger-Helmholtz
                      equation as numerical regularization of the nonlinear Schr\"{o}dinger equation}, Nonlinearity {\bf{21}} (2008) 879-898.


\bibitem{Cazenave} T. Cazenave,
               {\em Semilinear Schr\"{o}dinger Equations},
                Courant Lecture notes in Mathematics (2003), AMS.

\bibitem{Evans} L. C. Evans, {\em Partial Differential Equations}, Graduate Studies in Mathematics
                  vol {\bf{19}} (2000), AMS, Providence.

\bibitem{FibichDamp} G. Fibich, {\em Self-focusing in the damped nonlinear
                 Schr\"{o}dinger equation}, SIAM J. Appl. Math. {\bf{61}} (2001), 1680-1705.

\bibitem{FIP} G. Fibich, B. Ilan and G. Papanicolaou,
                 {\em Self-focusing with fourth-order dispersion},
                 SIAM J. on Appl. Math. {\bf 62} (2002), 1437-1462.


\bibitem{FibichLevy} G. Fibich and D. Levy,
                 {\em Self-focusing in the complex Ginzburg-Landau limit of the critical nonlinear
                    Schr\"{o}dinger equation}, Phys. Lett. A. {\bf{249}} (1998), 286-294.

\bibitem{FP1} G. Fibich and G. Papanicolaou, {\em A modulation method for self-focusing in the
                 perturbed critical nonlinear Schr\"{o}dinger equation}, Phys. Lett. A {\bf{ 239}} (1998),
                     167-173.


\bibitem{FP} G. Fibich and G. Papanicolaou,
               {\em Self-focusing in the perturbed and unperturbed nonlinear Schr\"{o}dinger
                 equation in critical dimenstion}, SIAM J. Appl. Math. {\bf{60}}, 183-240.


\bibitem{Velo} J. Ginibre and G. Velo,
               {\em On a class of nonlinear Schr\"{o}dinger equations. I. The Cauchy
                 problem, general case}, J. Funct. Anal. {\bf{32}} (1979), 1-32.

\bibitem{Glassey} R. T. Glassey,
                 {\em On the blowing-up of solutions to the Cauchy problem for the
                  nonlinear Schr\"{o}dinger equation},
                  J. Math. Phys. {\bf{18}} (1977), 1794-1797.

\bibitem{Kato} T. Kato, {\em On nonlinear Schr\"{o}dinger equations},
                           Ann. Inst. H. Poincar\'{e} Phys. Th\'{e}or. {\bf{46}} (1987), 113-129.


\bibitem{LST} E. Laedke, H. Spatschek and S. Turitsyn,
           {\em Analytics criterion for soliton instability in a nonlinear fiber array},
            Phys. Rev. E. {\bf{52}} (1995), 5549-5554.

\bibitem{Landman} M. Landman, G. Papanicolaou, C. Sulem and P. Sulem,
             {\em Rate of blowup for solutions of the nonlinear Schr\"{o}dinger equation
                 at critical dimension}, Phys. Rev. A {\bf{38}} (1988), 3837-3843.


\bibitem{Malkin} V. Malkin, {\em On the analytical theory for stationary self-focusing of radiation},
                  Physica D {\bf{64}} (1993), 251-266.


\bibitem{Sulem} C. Sulem and P. L. Sulem,
                {\em The Nonlinear Sch\"{o}dinger Equation,
                 Self-Focusing and Wave Collapse},
                 Applied Mathematical Sciences {\bf{139}} (1999), Springer-Verlag.


\bibitem{Weinstein} M. I. Weinstein,
               {\em Nonlinear Schr\"{o}dinger equations and sharp interpolation estimates},
                Commu. Math. Phys. {\bf{87}} (1983), 567-576.

\bibitem{WY} M. I. Weinstein and B. Yeary,
                {\em Excitation and dynamics of pulses in coupled fiber arrays},
                 Phys. Lett. A {\bf{222}} (1996), 157-162.


\end{thebibliography}
\end{document}